\documentclass[sigplan,nonacm]{acmart}

\setcopyright{none}
\renewcommand\footnotetextcopyrightpermission[1]{}
\acmConference[arXiv]{arXiv preprint}{June 2026}{}
\acmYear{2026}
\copyrightyear{2026}

\usepackage{amsmath}
\usepackage{listings}
\usepackage{algorithm}
\usepackage{algpseudocode}
\usepackage{xspace}
\usepackage{multirow}
\usepackage{placeins}
\usepackage{tikz}
\usepackage{tabularx}
\usepackage{subcaption}
\usepackage{enumitem}
\usepackage{microtype}

\newcommand{\sys}{SwarmX\xspace}

\newcommand{\mypar}[1]{\noindent\textbf{#1}\hspace{0.6em}}

\newcommand*\myc[1]{%
  \tikz[baseline=(c.base)]\node[shape=circle,fill=black,inner sep=.5pt,text=white](c){\scriptsize #1};%
}

\definecolor{codebg}{RGB}{253, 246, 227} 
\definecolor{codeborder}{RGB}{200, 200, 200}
\definecolor{codecomment}{RGB}{88, 110, 117} 
\definecolor{codestring}{RGB}{133, 153, 0}
\definecolor{codekeyword}{RGB}{38, 139, 210}

\definecolor{codeblue}{RGB}{0, 0, 255}       
\definecolor{codegreen}{RGB}{0, 150, 0}      
\definecolor{codeblack}{RGB}{0, 0, 0}        

\lstdefinestyle{Python}{
    language=Python,
    basicstyle=\ttfamily\fontsize{8pt}{8.5pt}\selectfont,
    keywordstyle=\color{codeblue},
    commentstyle=\color{codegreen},
    stringstyle=\color{codegreen},
    identifierstyle=\color{codeblack},
    numbers=left,
    numberstyle=\tiny\color{black}, 
    numbersep=6pt,                 
    frame=lines,                    
    rulecolor=\color{black},        
    showstringspaces=false,         
    tabsize=2,                      
    breaklines=true,                
    basewidth=0.5em,                
    fontadjust=true,                
    morekeywords={as, is, GNSPolicy},
    deletekeywords={None},
    showspaces=false,
    showstringspaces=false,
    captionpos=b, 
}

\newlength{\subfigH}
\begin{document}

\title{\sys: Agentic Scheduling for Low-Latency Agentic Systems}

\author{Yeqi Huang}
\affiliation{%
  \institution{University of Edinburgh}
  \country{United Kingdom}
}
\email{yeqi.huang@ed.ac.uk}

\author{Yanwei Ye}
\affiliation{%
  \institution{University of Edinburgh}
  \country{United Kingdom}
}
\email{yanwei.ye@ed.ac.uk}

\author{Guomin Chen}
\affiliation{%
  \institution{Tencent}
  \country{China}
}
\email{sarlmolchen@tencent.com}

\author{Wenhao Su}
\affiliation{%
  \institution{Tencent}
  \country{China}
}
\email{elonsu@tencent.com}

\author{Bin Gong}
\affiliation{%
  \institution{Tencent}
  \country{China}
}
\email{deargong@tencent.com}

\author{Jialian Li}
\affiliation{%
  \institution{Tencent}
  \country{China}
}
\email{jialianli@tencent.com}

\author{Zhan Lu}
\affiliation{%
  \institution{University of Edinburgh}
  \country{United Kingdom}
}
\email{zhan.lu@ed.ac.uk}

\author{Yangshen Deng}
\affiliation{%
  \institution{University of Edinburgh}
  \country{United Kingdom}
}
\email{yangshen.deng@ed.ac.uk}

\author{Xuan Sun}
\affiliation{%
  \institution{University of Edinburgh}
  \country{United Kingdom}
}
\email{xuan.sun@ed.ac.uk}

\author{Le Xu}
\affiliation{%
  \institution{University of Edinburgh}
  \country{United Kingdom}
}
\email{le.xu@ed.ac.uk}

\author{Luo Mai}
\affiliation{%
  \institution{University of Edinburgh}
  \country{United Kingdom}
}
\email{luo.mai@ed.ac.uk}

\renewcommand{\shortauthors}{Huang et al.}
\renewcommand{\shorttitle}{SwarmX: Agentic Scheduling for Low-Latency Agentic Systems}

\begin{abstract}
Agentic AI applications compose multiple model calls and tool executions, creating new scheduling challenges for GPU--CPU clusters. Their inference time and model-call structure often depend on prompt semantics, making conventional scheduling approaches ineffective for low-latency serving. This paper presents \sys{}, a system that implements \emph{agentic scheduling} for low-latency agentic applications. \sys{} uses scheduling-specific neural predictors to capture prompt, device, runtime, and target-model features; exposes distributional predictions to routers and scalers for tail-aware decisions; and provides mechanisms for predictor training and online adaptation. These predictors and mechanisms are integrated into a scheduler-agent framework that provides a common substrate for integration with existing scheduling and model-serving infrastructure. We evaluate \sys{} using production deployment (nearly one thousand GPUs and one million CPU cores) and controlled experiments on a 128-GPU testbed. Across multi-agent code generation, deep research, and multimodal agentic workflows, \sys{} reduces tail latency by up to \textbf{61.5\%} compared to state-of-the-art schedulers and sustains up to \textbf{2$\times$} the throughput of production schedulers under the same SLO\@.
\end{abstract}



\maketitle


\setlength{\textfloatsep}{8pt}
\setlength{\floatsep}{8pt}
\setlength{\abovecaptionskip}{2pt}
\setlength{\belowcaptionskip}{3pt}

\section{Introduction}
\label{sec:introduction}

Agentic AI is emerging as a new paradigm for building AI applications. Representative examples include multi-agent code generation~\cite{anthropic2025claudecode, codex2026}, synthetic data pipelines that coordinate multiple complementary models~\cite{wei2025deepseek, cui2025paddleocr30technicalreport, liu2025toolace, zhou2024staragents}, and multimodal generative applications built around agentic workflows~\cite{leviathan2025generativeui, comfyui2026}. These applications typically run on GPU--CPU clusters, where each model is deployed as a replicated service across GPUs or CPUs. A router, or request scheduler, selects the replica that serves each request, while a scaler, or resource scheduler, decides how many replicas each model should run. The goal is to minimize the end-to-end latency of composed multi-model inference while maintaining high GPU and CPU utilization.

Scaling agentic AI on GPU--CPU clusters is challenging because the workload behavior is both dynamic and prompt-dependent. First, a single model call may require a long and highly variable number of decoding steps, depending on the semantics of the input prompt. We call this challenge \emph{prompt-dependent variable inference time}. Second, agentic applications often exhibit dynamic model-to-model calling patterns. The downstream model to invoke may be determined only at runtime, after an upstream model produces its output. We call this challenge \emph{prompt-dependent model-call structure}. Together, these two forms of dynamism make both routing and scaling decisions difficult: the scheduler must decide where to place a request and how much capacity to provision before fully knowing how long the request will run or which downstream models it will trigger.

Existing schedulers fall short in addressing these challenges. De-facto AI cluster schedulers such as Ray~\cite{ray} and Kubernetes~\cite{kubernetes} often rely on simple policies, including round-robin, random scheduling, or power-of-two choices~\cite{mitzenmacher2002power}. These policies work well when QPS is high and requests are short, because poor decisions are quickly amortized. Agentic workloads violate this assumption: requests can run much longer through an agent harness, so poor routing or scaling decisions persist and amplify tail latency. Predictor-based schedulers such as PSC~\cite{faisal2024will} and Cilantro~\cite{288542} use learned models, but typically rely on simple predictors, such as linear regression or random forests, which cannot accurately capture prompt-dependent inference time or model-call structure. Recent agentic schedulers, including ORION~\cite{orion}, Pie~\cite{gim2025pie}, and Parrot~\cite{lin2024parrot}, mainly target single-agent or single-workload settings. Murakkab~\cite{murakkab} supports multi-model scheduling, but relies on average per-model estimates and ignores prompt semantics, leaving it vulnerable to poor latency performance.

In this paper, we explore a new approach to scheduling agentic AI applications on GPU--CPU clusters. Our key insight is that, because both inference time and model-call structure depend on prompt semantics, schedulers need \emph{scheduling-specific neural predictors} that combine prompt semantics with device, runtime, and target-model characteristics. These predictors allow schedulers to estimate request latency and downstream resource demand before making routing and scaling decisions. However, using neural predictors for scheduling is not simply a matter of adding a model. A practical system must manage predictors online, keep prediction overhead low, translate distributional predictions into concrete routing and scaling actions, and coordinate these actions across model services to optimize end-to-end latency and cluster-wide resource utilization.

We realize this insight in \sys{}, a system that implements \emph{agentic scheduling} for serving agentic AI applications on GPU--CPU clusters. In \sys{}, routers and scalers become lightweight scheduler agents: they observe prompt, target-model, device, and runtime state; use neural predictors to anticipate latency and model-call behavior; and take bounded routing or scaling actions through existing scheduler interfaces. This augments existing cluster managers with prediction-driven, uncertainty-aware control without replacing their infrastructure.

\sys{} makes the following contributions.

\mypar{(1)~Prompt-, device-, runtime-, and target-model-aware predictors.}
We design compact neural predictors for agentic AI scheduling. These predictors capture prompt semantics, device characteristics, runtime state, and target-model properties to estimate inference latency and downstream model-call behavior with low deployment overhead.

\mypar{(2)~Distribution-aware scheduling.}
\sys{} exposes distributional predictions directly to routers and scalers instead of reducing them to point estimates. It composes these distributions across sequential routing and scaling decisions, allowing schedulers to preserve predictive uncertainty and control tail latency.

\mypar{(3)~Training and adaptation of predictors.}
\sys{} provides mechanisms for training predictor components with router- and scaler-specific objectives, monitoring prediction quality online, and retraining lightweight components when workload shifts degrade accuracy.

\mypar{(4)~Scheduler-agent framework for integration.}
\sys{} embeds neural prediction into existing schedulers through a scheduler-agent framework. The framework represents scheduling operations as bounded agent actions and manages the data used to train, monitor, and adapt predictors as agent memory, enabling low-cost integration with existing scheduling and model-serving infrastructure.

We implement \sys{} as a scheduler plug-in for existing AI cluster infrastructure. The implementation is designed for scalability, reliability, low overhead, and compatibility with existing programming interfaces. \sys{} has been deployed in production clusters and used to support a range of emerging agentic AI applications.

We report deployment results from one of our production clusters, which contains several million CPU cores and nearly one thousand heterogeneous GPUs, and complement them with controlled experiments on a 128-GPU testbed.
Across important agentic workloads---multi-agent code generation (Coding Agent\cite{codex2026,opencode2026},
OpenClaw\cite{openclaw2026}), deep research\cite{du2025deepresearch}, and multimodal agentic workflows such as text-to-video\cite{comfyui2026}
generation---\sys{} outperforms Ray~\cite{ray}, Murakkab~\cite{murakkab},
Power-of-Two Choices~\cite{mitzenmacher2002power}, and our production schedulers.
In production, \sys{} sustains up to \textbf{2$\times$} the throughput of the prior scheduler under the same SLO and reduces P99 latency by \textbf{44--52\%}. On the controled testbed, it reduces end-to-end P95 latency up to \textbf{61.5\%} even on open-ended agentic workloads whose call structure is decided entirely at runtime.

\section{Background and Motivation}
\label{sec:background}

\subsection{Agentic applications on GPU--CPU clusters}
\label{sec:background:cluster}

\begin{figure}[t]
\centering
\includegraphics[width=0.9\columnwidth]{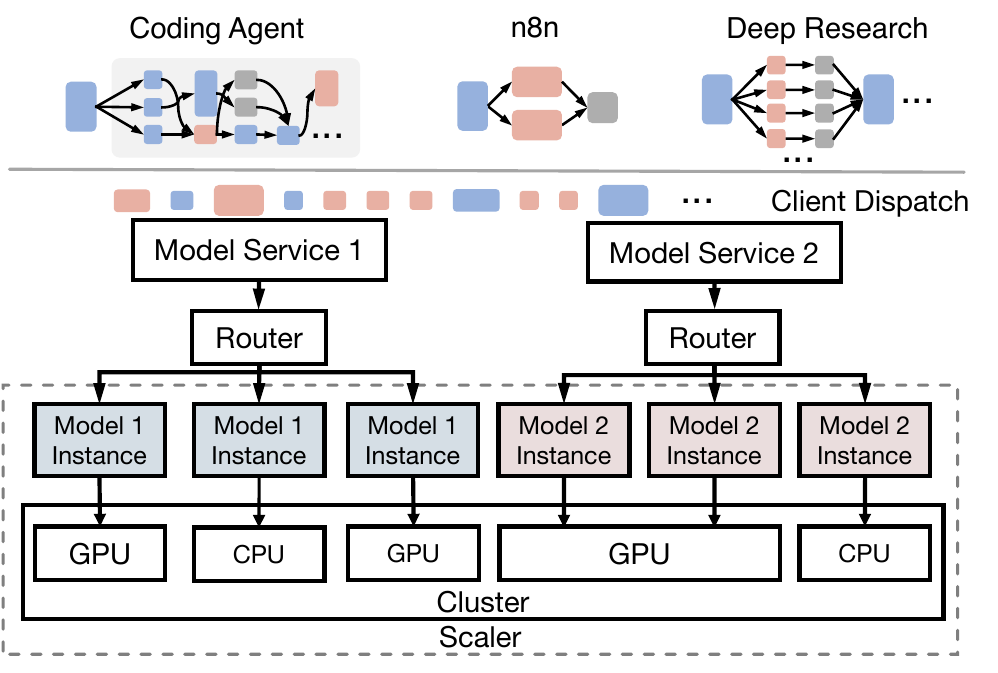}
\caption{Overview of a typical agentic application cluster.}
\Description{Diagram showing a typical agentic application cluster, with users initiating workflows that issue many model calls served by replicated model-as-a-service backends behind routers and scalers.}
\label{fig:traditional_system_overview}
\end{figure}

An agentic application handles a user request by repeatedly invoking AI models and tools in response to a prompt. Examples include OpenClaw-style agent workloads, coding agents, data-cleaning and synthetic-data generation pipelines, and interactive media-generation workflows such as ComfyUI.

A key property of these applications is that execution spans multiple models, agents, or tool calls. As shown in \autoref{fig:traditional_system_overview} top, a coding agent may alternate between planning, code search, editing, testing, debugging, and summarization; each stage may invoke different models, prompts, or tools. Similarly, an OpenClaw-style workload may contain many coordinated agent steps whose dependencies are only partially known before execution. This makes the serving target fundamentally different from optimizing a single model.

Serving such applications often requires large GPU--CPU clusters. As shown in \autoref{fig:traditional_system_overview} bottom, each model is typically deployed as an independent service, such as a model-as-a-service endpoint or a Ray actor/task interface. To sustain throughput, each service is backed by multiple replicas. Two scheduling components govern service performance: (i)~\emph{routers}, which dispatch requests to replicas to balance load and reduce latency, and (ii)~\emph{scalers}, which adjust replica counts and placements across GPUs and CPUs to avoid under- or over-provisioning.


\newlength{\legendH}
\newsavebox{\rightCol}

\begin{figure*}[t]
    \centering
    \begin{minipage}[b]{0.3\linewidth}
        \centering
        \rule{0pt}{0pt}\vspace{0pt}\par
        \includegraphics[width=\linewidth]{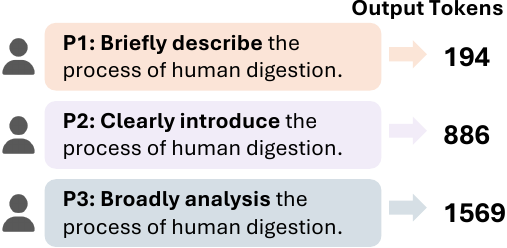}\\[10pt]
        \subcaption{Prompt and output tokens.}
        \label{fig:prompt-output-token}
    \end{minipage}\hspace{12pt}%
    \begin{minipage}[b]{0.6\linewidth}
        \centering
        \begin{minipage}[b]{0.49\linewidth}
            \centering
            \includegraphics[width=\linewidth]{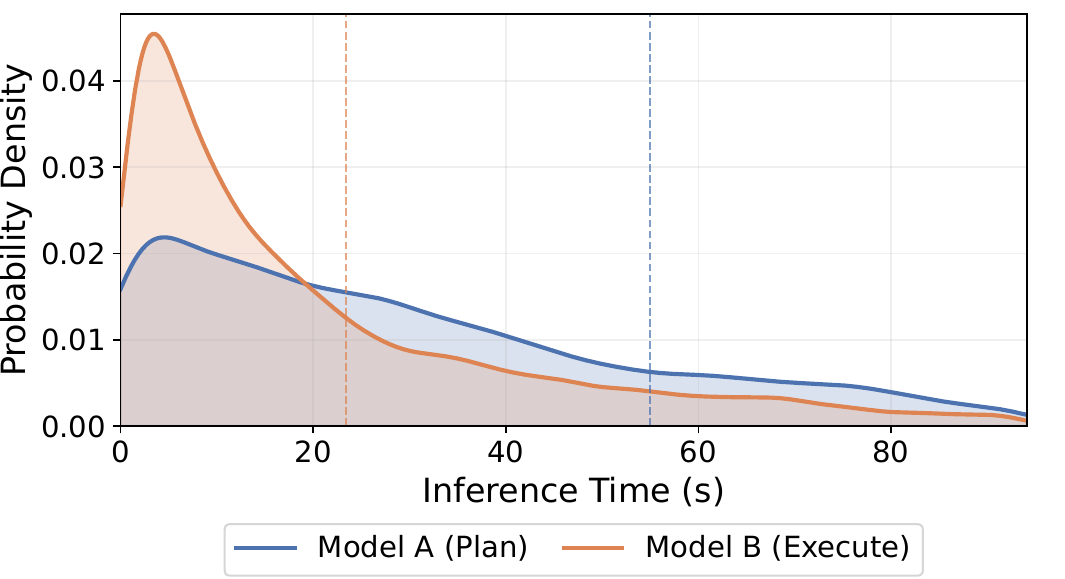}
        \end{minipage}%
        \hfill
        \begin{minipage}[b]{0.49\linewidth}
            \centering
            \includegraphics[width=\linewidth]{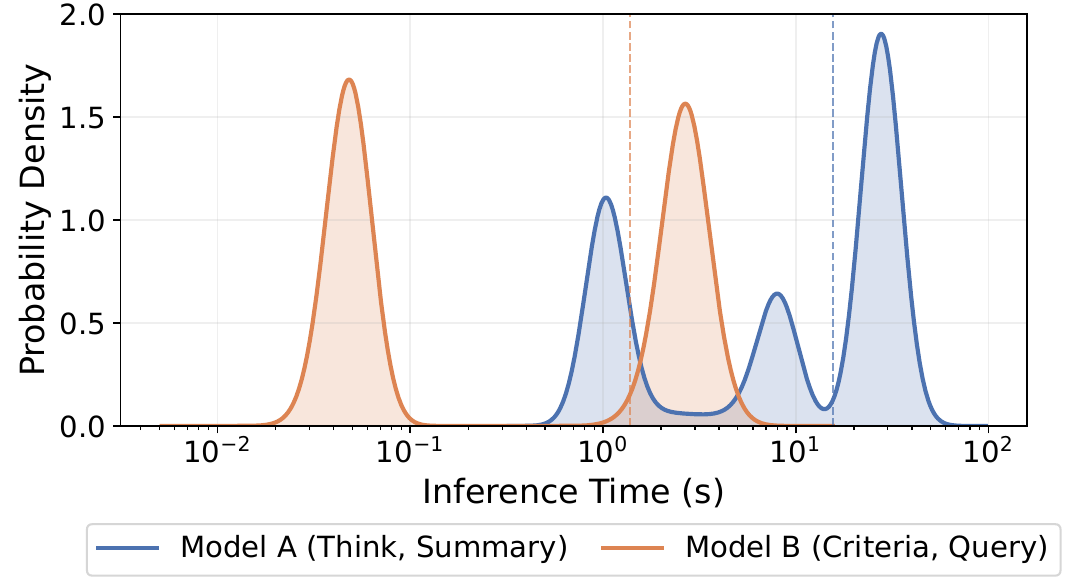}
        \end{minipage}\\[0pt]
        \begin{subfigure}[b]{0.49\linewidth}
            \subcaption{Coding Agent.}
            \label{fig:inference-time-cc}
        \end{subfigure}%
        \hfill
        \begin{subfigure}[b]{0.49\linewidth}
            \subcaption{Deep Research.}
            \label{fig:inference-time-dr}
        \end{subfigure}
    \end{minipage}
    \caption{Inference time depends strongly on prompt semantics, and its distribution varies across both models and workloads.}
    \label{fig:prompt-token-inference-time}
\end{figure*}

\subsection{Scheduling Challenges for Agentic Systems}
\label{sec:background:challenges}

Agentic systems invoke large language models and multimodal models whose requests may take seconds to minutes to complete. This makes each routing decision consequential and requires scalers to provision replicas before queues build up. Poor decisions, such as placing a long request behind an already slow queue or failing to scale replicas in time, can significantly increase tail latency and reduce resource utilization. Agentic systems therefore require routers and scalers to make consistently high-quality decisions under two forms of prompt-dependent dynamism.

\mypar{(1)~Prompt-dependent variable inference time.}
Inference time depends strongly on the input prompt. As shown in Figure~\ref{fig:prompt-token-inference-time}(a), three semantically similar prompts generate vastly different numbers of output tokens (194, 886, and 1569), leading to substantially different inference times. This variability also differs across models within the same workload (Figure~\ref{fig:prompt-token-inference-time}(b)) and across workloads (Figure~\ref{fig:prompt-token-inference-time}(c) vs.\ (b)).

This creates a direct challenge for routers. When assigning requests to model-replica queues, a router needs to estimate how quickly each queue will drain. However, without prompt-aware prediction, requests that appear similar may occupy replicas for very different lengths of time. This makes load balancing unreliable and tail latency difficult to protect.

\begin{figure*}[t]
    \centering
    \begin{minipage}[b]{0.35\linewidth}
        \centering
        \includegraphics[width=0.88\linewidth]{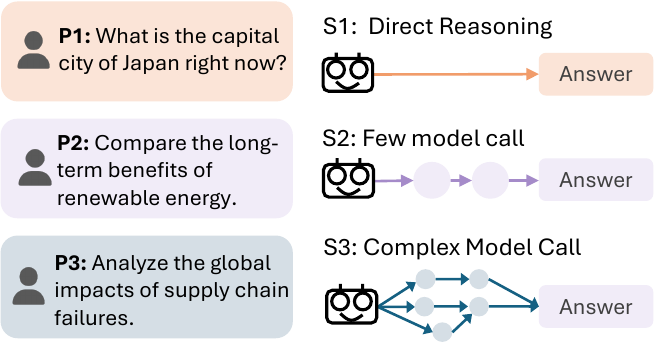}\\[0pt]
        \subcaption{Prompt and its following model-call structure.}
        \label{fig:dynamic_model_call}
    \end{minipage}\hspace{12pt}%
    \begin{minipage}[b]{0.6\linewidth}
        \centering
        \begin{minipage}[b]{0.49\linewidth}
            \centering
            \includegraphics[width=\linewidth]{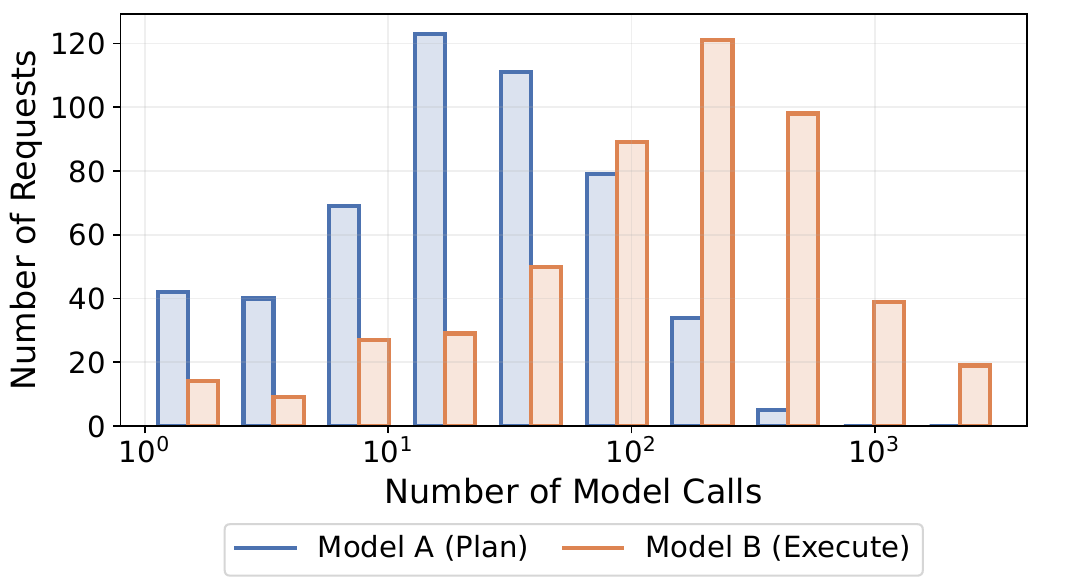}
        \end{minipage}%
        \hfill
        \begin{minipage}[b]{0.49\linewidth}
            \centering
            \includegraphics[width=\linewidth]{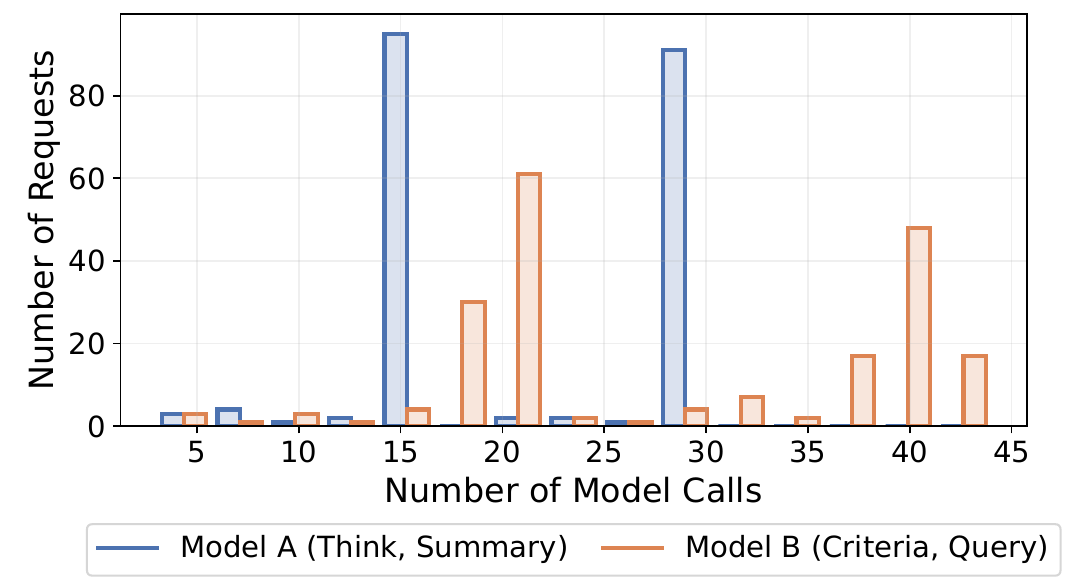}
        \end{minipage}\\[0pt]
        \begin{subfigure}[b]{0.49\linewidth}
            \subcaption{Coding Agent.}
            \label{fig:model_calls_cc}
        \end{subfigure}%
        \hfill
        \begin{subfigure}[b]{0.49\linewidth}
            \subcaption{Deep Research.}
            \label{fig:model_calls_dr}
        \end{subfigure}
    \end{minipage}
    \caption{The number and structure of model calls are prompt-dependent, and their distribution varies across both models and workloads.}
    \label{fig:prompt-token-model-call-structure}
\end{figure*}

\mypar{(2)~Prompt-dependent model-call structure.}
Agentic applications issue multiple model calls through an agent harness, and both the number and structure of these calls depend on the prompt. As shown in Figure~\ref{fig:prompt-token-model-call-structure}(a), prompts of increasing complexity induce distinct execution structures: a direct one-call answer (S1), a short chain (S2), and a complex DAG (S3). The number-of-calls distribution also differs across models within a workload (Figure~\ref{fig:prompt-token-model-call-structure}(b)) and across workloads (Figure~\ref{fig:prompt-token-model-call-structure}(c) vs.\ (b)).

This creates a direct challenge for scalers. A scaler must decide how many replicas to provision for each model, but the required capacity depends on model-call structures that are revealed only at runtime. As a result, scalers may under-provision downstream models, over-provision unused replicas, or react too late to shifting demand, all of which hurt tail latency and resource utilization.

\subsection{Limitations of Existing Scheduling Approaches}

To handle scheduling dynamism in agentic systems, one could consider three categories of existing approaches.

\mypar{(1)~De-facto AI cluster scheduling.}
Ray and Kubernetes are widely used in AI clusters, but their scheduling decisions often rely on simple policies, such as round-robin, random scheduling, or power-of-two choices~\cite{mitzenmacher2002power}. These policies work well when requests are short and QPS is high, because poor decisions are quickly amortized. Agentic workloads violate this assumption: requests can run much longer through an agent harness, while QPS is often lower. As a result, a poor routing or scaling decision can persist longer, causing queue imbalance and amplifying tail latency.

\mypar{(2)~Scheduling with learned or statistical predictors.}
Another approach is to predict request cost using learned or statistical models. Systems such as PSC~\cite{faisal2024will} and Cilantro~\cite{288542} follow this direction. However, they typically rely on simple predictors, such as linear regression or random forests, which are insufficient for capturing prompt-dependent inference time and model-call structure in agentic workloads. They also do not address the system requirements of using more capable neural predictors, including dataset construction, low-overhead deployment, online adaptation, and integration with router and scaler decisions.

\mypar{(3)~Scheduling systems for agentic workloads.}
Recent systems have begun to target agentic workloads directly. ORION~\cite{orion}, Pie~\cite{gim2025pie}, and Parrot~\cite{lin2024parrot} are representative examples, but they primarily focus on single-agent or single-workload settings rather than the multi-agent, multi-model workloads studied in this paper. Murakkab~\cite{murakkab} supports multi-model scheduling, but estimates per-model inference time using average values and remains unaware of prompt semantics. Pythia~\cite{yu2026pythiaexploitingworkflowpredictability} is a parallel effort that predicts workflow structure in multi-agent workflows, but does not model prompt semantics. As a result, these systems can still make poor tail-latency decisions when prompt-dependent inference time is highly variable, as shown in Figure~\ref{fig:prompt-token-inference-time}.

\section{\sys{} Design}
\label{sec:scheduler_agent}

\begin{figure}[t]
    \centering
    \includegraphics[width=0.9\linewidth]{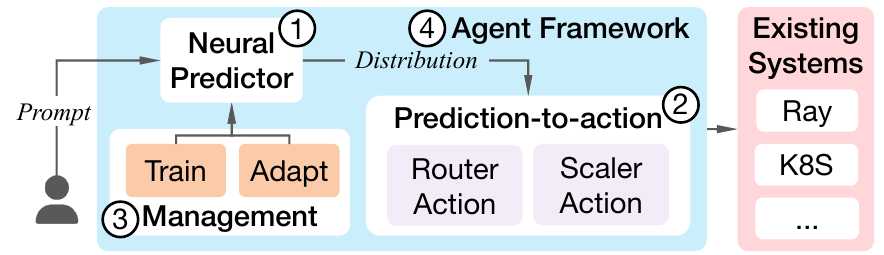}
    \caption{\sys{} design overview.}
    \label{fig:agent_design}
\end{figure}

We design \sys{} around four goals:
\begin{itemize}[leftmargin=1.2em,itemsep=0.2em,topsep=0.2em]
    \item \textbf{G1: Making prediction aware of prompt, device, runtime, and target model.}
    \sys{} should capture how prompt semantics, device characteristics, runtime conditions, and target-model properties affect inference time and model-call structure.

    \item \textbf{G2: Ensuring robust routing and scaling decisions based on prediction distributions.}
    \sys{} should translate the potentially complex distributions of predictions made by neural networks into concrete routing and scaling actions, respecting the different objectives of routers and scalers and remaining general across scheduling scenarios.

    \item \textbf{G3: Unifying training and adaptation management for predictors.}
    \sys{} should provide common mechanisms for training predictors, monitoring prediction quality, and triggering retraining when prediction quality degrades.

    \item \textbf{G4: Ensuring low-cost integration with existing infrastructure.}
    \sys{} should integrate with existing cluster infrastructure through clean interfaces for reading runtime and device state and invoking bounded scheduling actions available to routers and scalers.
\end{itemize}

\mypar{Design idea and overview.}
Figure~\ref{fig:agent_design} shows the design of \sys{}. The core idea is to turn routers and scalers into lightweight scheduler agents. Each agent observes prompt, target-model, device, and runtime state; invokes neural predictors to estimate inference-time and model-call distributions; and translates these predictions into bounded routing or scaling actions exposed by existing infrastructure, such as Ray or Kubernetes. We call this design \emph{agentic scheduling}.

\sys{} realizes agentic scheduling through four components. A \emph{prompt-, device-, runtime-, and target-model-aware predictor} (\myc{1}, Section~\ref{sec:NN}) estimates inference-time and model-call distributions. A \emph{distribution-aware prediction-to-action mechanism} (\myc{2}, Section~\ref{sec:quantile}) converts these distributions into routing and scaling actions while preserving uncertainty. \emph{Shared training and adaptation mechanisms} (\myc{3}, Section~\ref{sec:memory}) monitor prediction quality and retrain predictors under workload shifts. Finally, a \emph{scheduler-agent framework} (\myc{4}, Section~\ref{sec:interface}) wraps predictors, data, adaptation logic, and action interfaces into a common substrate for integration with existing routers, scalers, and cluster managers.

\subsection{Predictor Design}
\label{sec:NN}

\sys{} uses neural predictors that combine prompt semantics with device, runtime, and target-model information to support routing and scaling decisions.

\mypar{Decouple neural components for different forms of awareness.}
A predictor must incorporate a wide range of features that affect scheduling decisions. Prompt semantics are especially challenging: interpreting a prompt typically requires a language model, which can be expensive to train and run. Our key observation is that prompt understanding does not need to be learned from scratch. Smaller variants of the target model, which we call \emph{isomorphic small variants}, already capture useful semantic signals about the difficulty and likely structure of the response generated by the target model.

\begin{figure}[t]
    \centering
    \includegraphics[width=0.8\linewidth]{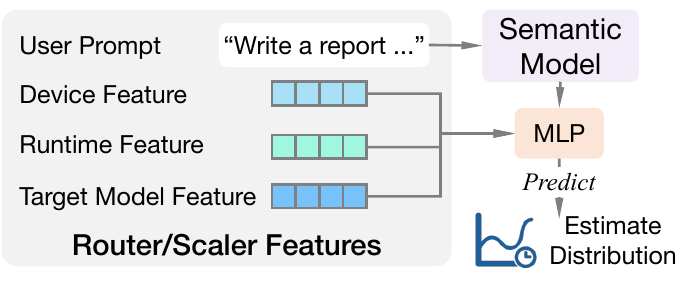}
    \caption{Design overview of the neural predictor.}
    \label{fig:neural-network-overview}
\end{figure}

Motivated by this observation, we decouple the predictor into two components, as shown in Figure~\ref{fig:neural-network-overview}.

\begin{itemize}[leftmargin=1.2em,itemsep=0.2em,topsep=0pt,parsep=0pt,partopsep=0pt]
    \item \emph{A semantic model} aligned with the target model being predicted. It extracts semantic features from the prompt and can be initialized from an existing pre-trained language model, reducing training cost. Architecturally, it is a smaller isomorphic variant of the target model, with fewer layers or fewer parameters per layer.

    \item \emph{A multi-layer perceptron (MLP)} that combines semantic features with device, runtime, and target-model features. The MLP produces a distributional prediction that can be consumed by downstream routers and scalers.
\end{itemize}

\mypar{Router- and scaler-oriented prediction.}
The feature set depends on the scheduling objective and the information observable in the deployment environment.

For router-oriented prediction, \sys{} uses four groups of input features: (i)~semantic features extracted from the user prompt, which capture the likely difficulty and response structure of the request; (ii)~device features, such as hardware type, available compute cores, clock frequency, and theoretical FLOPS; (iii)~runtime features, including utilization, active concurrency, inference runtime version, and inference-engine parameters such as maximum batch size; and (iv)~target-model features extracted from the model configuration, such as hidden size and number of layers. The predictor outputs a latency distribution for the request, enabling the router to make tail-aware placement decisions rather than relying on a point estimate.

For scaler-oriented prediction, \sys{} uses a more compact feature set: (i)~the same semantic features used by the router, (ii)~device features that describe the hardware types in the current deployment, and (iii)~runtime features that describe the current replica list and each replica's state. The predictor outputs downstream model-call distributions.

\mypar{Keep prediction overhead low.}
This decoupled design keeps the predictor orders of magnitude smaller than the target models used by agentic applications. First, the relationship between device/runtime state and latency or demand can be captured by a lightweight MLP. Second, predicting latency or downstream model-call structure is much simpler than performing the target model's reasoning task, so semantic features can be extracted by a much smaller language model. As a result, the predictor can run on CPUs or consume only a small fraction of GPU resources. This keeps prediction overhead low, as confirmed by our evaluation and production deployment experience.

\mypar{Searching for model size.}
We formulate model sizing as an accuracy--overhead search problem. The MLP is intentionally small, typically with hundreds of thousands to at most a few million parameters. The semantic model is larger, but still several orders of magnitude smaller than the target model. For example, for an 8B-parameter Qwen target model, we select a 35M-parameter semantic model. Our evaluation shows that this size preserves prediction accuracy while keeping prediction overhead low.

\subsection{Prediction-Distribution-Aware Scheduling}
\label{sec:quantile}

\begin{figure}[t]
    \centering
    \includegraphics[width=1.0\linewidth]{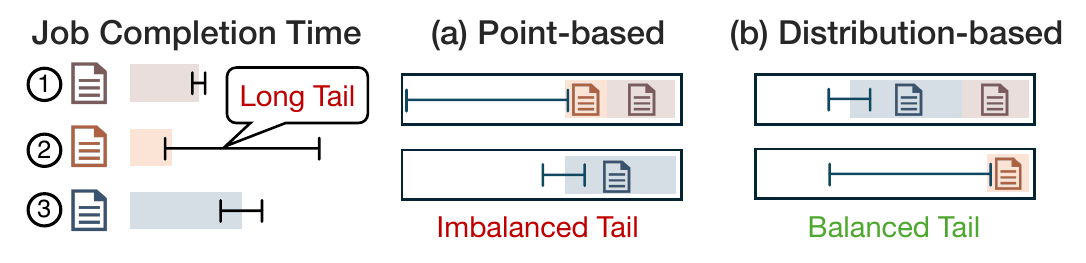}
    \caption{Point-based composition can hide tail behavior; distribution-based composition preserves tail awareness.}
    \label{fig:distribution_scheduling}
\end{figure}

\sys{} turns neural predictions into routing and scaling decisions through distribution-aware scheduling.

\mypar{Insight: Prediction distributions preserve scheduling-relevant information that point estimates discard.}
Conventional schedulers often act on a single predicted value, such as mean latency. This point estimate can hide skewed, heavy-tailed, or multi-modal behavior, which is common in LLM serving because latency depends strongly on prompt semantics. As shown in \autoref{fig:distribution_scheduling}, compact and long-tailed distributions may have similar point estimates but very different tail risks. After composition, point estimates also obscure how individual tails contribute to the final outcome. \sys{} therefore keeps predictions distributional until the final scheduling decision, which is especially important for low-QPS agentic workloads where one poor decision can dominate tail latency.

\mypar{Represent uncertainty with quantiles.}
\sys{} represents both predicted distributions and maintained scheduling state with quantile sketches. Quantiles preserve distribution shape and tail behavior while remaining cheap to store, compose, and update online. This allows each new prediction to be combined incrementally with uncertainty already accumulated in queues or demand states.

\mypar{Distribution-composition template.}
Scheduling decisions are sequential: each routing or scaling action changes queue state or replica demand, which affects later decisions. \sys{} therefore maintains a compact distributional state $\mathbf{S}$ whose entries summarize committed work, such as per-queue completion sketches for routers or demand sketches for scalers. For each candidate action $a$, \sys{} predicts the induced work distribution $D_a$ and constructs a hypothetical state $\mathbf{S}^{(a)}$ by composing $D_a$ only with the affected entry using quantile-grid composition $\oplus$. The scheduler then applies a cost evaluator $\mathcal{C}$ to the whole candidate state, producing a distributional cost $c_a=\mathcal{C}(\mathbf{S}^{(a)})$ for objectives such as P99 latency or throughput loss. Since $c_a$ remains distributional, \sys{} samples candidate actions from the distribution induced by $\{c_a\}$, selects the best sampled candidate, and commits only the selected state update.

\begin{algorithm}[t]
\caption{Distribution-Aware Request Routing}
\label{alg:route}
\small
\begin{algorithmic}[1]
\Procedure{Route}{$r, G, \mathbf{Q}$}
\Statex \hspace{\algorithmicindent}\textit{Inputs: predictor $F$; device features $\tau$; runtime features $\sigma$}
\Statex \hspace{\algorithmicindent}\textit{Tail-cost evaluator $\mathcal{C}_{\mathrm{tail}}$; queue sketches $\mathbf{Q}$}
\ForAll{$g \in G$}
    \State \label{alg:route:predict} $D_g \gets F\!\left(r,\, \tau(g),\, \sigma(g)\right)$
        \Comment{Predicted latency distribution}
    \State \label{alg:route:compose} $\mathbf{Q}[g] \gets \mathbf{Q}[g]\oplus D_g$
        \Comment{Update queue $g$}
    \State \label{alg:route:score} $c_g \gets \mathcal{C}_{\mathrm{tail}}\!\left(\mathbf{Q}\right)$
        \Comment{Tail-cost distribution}
\EndFor
\State \label{alg:route:sample} $\mathcal{S} \gets \Call{Sample}{\{c_g\}}$
    \Comment{Probability-aware subset}
\State \label{alg:route:draw} $\hat{c}_g \sim c_g\ (g\in\mathcal{S})$
    \Comment{Sample tail costs}
\State \label{alg:route:select} $g^\star \gets \arg\min_{g \in \mathcal{S}}\, \hat{c}_g$
\State \label{alg:route:dispatch} \Call{Dispatch}{$r, g^\star$}
    \Comment{Send request to selected queue}
\State \Return $g^\star$
\EndProcedure
\end{algorithmic}
\end{algorithm}

\mypar{Instantiating the template in routers and scalers.}
Algorithm~\ref{alg:route} instantiates the template for request routing. The generic state $\mathbf{S}$ becomes a vector of queue sketches $\mathbf{Q}$, and each candidate action becomes a queue $g\in G$. For each candidate, \sys{} predicts the request-latency distribution from the request, device features, and runtime features (Line~\ref{alg:route:predict}), and composes that distribution only into the queue-$g$ sketch to form the candidate state (Line~\ref{alg:route:compose}). The tail-cost evaluator $\mathcal{C}_{\mathrm{tail}}$ is then applied to the full queue state (Line~\ref{alg:route:score}), so a single-entry update is still judged by its effect on the whole schedule before the selected queue is dispatched (Lines~\ref{alg:route:sample}--\ref{alg:route:dispatch}).

Scalers use the same instantiation pattern with demand sketches in place of queue sketches and candidate scaling decisions or target deployments in place of candidate queues. At each scaling interval, \sys{} scores hypothetical demand states and commits the sampled best candidate; the implementation also applies a deployment-change threshold to avoid reacting to small demand fluctuations.

\subsection{Training and Adaptation of Predictors}
\label{sec:memory}

\sys{} trains its predictors from production traces and adapts them when online prediction quality degrades.

\mypar{Training method design.}
The key design choice is the loss function for each predictor component. \sys{} uses separate objectives for the semantic model and the MLPs used by routers and scalers.

First, the semantic model is trained to predict prompt-level properties of the target model, such as output token length or response-structure features:
\begin{equation}
    \mathcal{L}_{\mathrm{sem}}
    =
    \frac{1}{|\mathcal{D}_{\mathrm{sem}}|}
    \sum_{(p,\,m,\,y) \in \mathcal{D}_{\mathrm{sem}}}
    \rho\bigl(\hat{y}_{p,m},\, y\bigr),
\end{equation}
where $p$ is the input prompt, $m$ is the target model, $y$ is the observed target-model response property, $\hat{y}_{p,m}$ is the semantic model's prediction, and $\rho(\cdot,\cdot)$ is a configurable per-sample loss, such as MSE, MAE, Huber, or pinball loss.

Second, the router MLP is trained to predict an inference-time distribution using weighted pinball loss over prescribed quantile levels:
\begin{equation}
    \mathcal{L}_{\mathrm{router}}
    =
    \frac{1}{|\mathcal{D}_{\mathrm{router}}|}
    \sum_{(x,\,t) \in \mathcal{D}_{\mathrm{router}}}
    \sum_{k=1}^{K}
    w_k \,
    \rho_{\tau_k}\!\left(t - \hat{q}_{\tau_k}(x)\right),
\label{eq:pinball_loss}
\end{equation}
where $x$ contains semantic, target-model, device, and runtime features; $t$ is the observed inference time; $\hat{q}_{\tau_k}(x)$ is the predicted $\tau_k$-quantile; $w_k>0$ and $\sum_{k=1}^{K}w_k=1$; and $\rho_{\tau}(u)=\max(\tau u,(\tau-1)u)$ is the standard pinball loss.

The scaler MLP uses the same weighted pinball-loss form, but applies it across the predicted downstream call-count distributions for all target models. This trains the scaler to preserve uncertainty over future model demand without introducing a separate objective.

We construct the training dataset from logs collected during application execution. Each record contains the prompt context, target-model information, device and runtime features, prediction output, scheduling decision, and observed outcome. We train each predictor component until convergence using standard deep learning optimizers, such as AdamW.

\begin{algorithm}[t]
\caption{Online Adaptation}
\label{alg:ood}
\small
\begin{algorithmic}[1]
\Procedure{Adapt}{$p, g, D_p, \ell$}
\Statex \hspace{\algorithmicindent}\textit{Inputs: windows $W$; capacity $N$;}
\Statex \hspace{\algorithmicindent}\textit{threshold $\theta$; tail level $\alpha$}
\State $k \gets \mathrm{key}(p,\, \tau(g))$
    \Comment{Prompt/device group}
\State $e \gets \rho_\alpha\!\left(\ell - \mathcal{Q}_\alpha(D_p)\right)$
    \Comment{Tail pinball loss}
\State \Call{Push}{$W_k, e, N$}
    \Comment{Maintain sliding window}
\If{$\frac{1}{|W_k|}\sum_{e' \in W_k} e' > \theta$}
    \State \Call{RetrainMLP}{$k$}
        \Comment{Async retraining}
    \State $W_k \gets \emptyset$
\EndIf
\EndProcedure
\end{algorithmic}
\end{algorithm}

\mypar{Predictor online adaptation.}
Our adaptation method is guided by one key observation: most online drift comes from changes in workload mix, system load, device utilization, or runtime behavior. These shifts affect the mapping from semantic, target-model, device, and runtime features to latency or demand, but usually do not change the target model's prompt-to-response behavior. In such cases, retraining the lightweight MLP is sufficient. The semantic model is retrained only after target-model changes or updates.

Based on this observation, we design Algorithm~\ref{alg:ood}, which defines how \sys monitors prediction quality and triggers retraining under distribution shift. It treats a predictor as out-of-distribution (OOD) when recent observed latencies deviate from the predicted distribution beyond a threshold. When a request completes, \sys{} groups it by prompt class and device type, computes the tail pinball loss between the observed latency and the predicted tail quantile, and appends the error to the corresponding sliding window. If the average window error exceeds the configurable threshold $\theta$, Algorithm~\ref{alg:ood} asynchronously retrains the corresponding MLP using recent records from the same window. Online scheduling continues with the current predictor during retraining. Once the retrained MLP passes validation, \sys{} installs it for subsequent scheduling.

\subsection{Scheduler-Agent Framework as a Substrate}
\label{sec:interface}

\begin{figure}[t]
    \centering
    \includegraphics[width=0.8\linewidth]{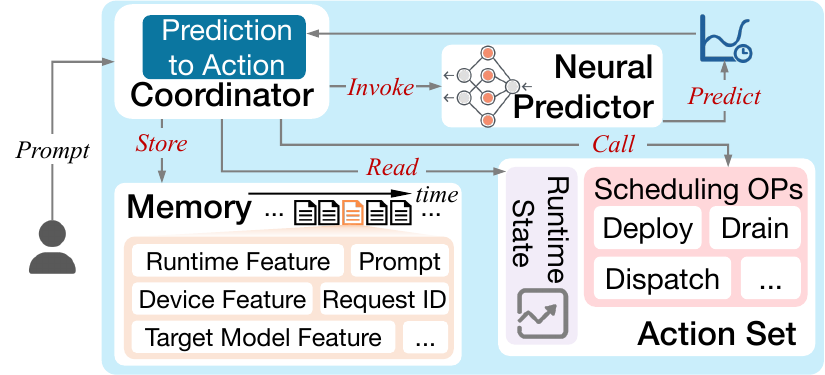}
    \caption{\sys{} scheduler-agent framework.}
    \label{fig:agent_framework}
\end{figure}

\sys{} introduces a scheduler-agent framework that integrates neural predictors, prediction data, adaptation logic, and scheduling actions into a common substrate. The framework turns routers and scalers into lightweight scheduler agents while preserving clean interfaces to existing scheduling and model-serving infrastructure.

Figure~\ref{fig:agent_framework} shows this boundary. \texttt{Predictor}, \texttt{Coordinator}, and \texttt{Memory} capture the reusable logic described in the previous subsections: prediction, distribution-aware decision making, and online adaptation. The \texttt{Action Set} is the infrastructure-specific boundary, exposing only the runtime-state reads and scheduling operations that an agent is allowed to use.

\mypar{Clean interfaces to existing scheduling infrastructure.}
The \texttt{Action Set} exposes two classes of primitives: runtime-state reads, such as GPU utilization and concurrency; and bounded scheduling operations, such as dispatching a request (\texttt{Dispatch}), adding a model replica (\texttt{Deploy}), and removing a model replica (\texttt{Drain}). The \texttt{Coordinator} can act only through these primitives, so every agent action is mediated by a controlled interface that preserves system safety and stability. Different agents can bind different \texttt{Action Set}s while reusing the same predictor, adaptation, and distribution-aware decision logic.

\mypar{Coordination between scheduler agents.}
Scheduler agents coordinate by exchanging compact state-change notifications rather than centralizing every decision. For example, when a scaler deploys or drains replicas, it publishes the updated replica set to affected routers; routers then refresh candidate queues and local runtime state before making subsequent routing decisions. Routers therefore make routing decisions without waiting for the scaler, but still update their candidate queues promptly when replicas are added or removed.

\section{\sys{} Implementation}
\label{sec:implementation}

\sys{} is implemented as a scheduler plug-in for existing AI cluster infrastructure. The implementation targets scalability, low overhead, reliability, and compatibility with existing workload and application interfaces.

\mypar{System architecture.}
Worker servers execute AI and agentic models using inference engines such as SGLang\cite{sglang} and vLLM\cite{vllm}, and report lightweight heartbeats to the control plane. Workers are grouped by application to reduce performance interference. A cluster-wide resource coordinator manages elasticity and cross-application resource arbitration.

Control servers manage workers and provide service discovery, allowing clients to locate required model services dynamically. For each workflow, \sys{} deploys multiple control servers using a shared-state scheduling design~\cite{omega}, improving reliability and scalability without changing existing application interfaces.

\mypar{Deploying neural predictors.}
\sys{} triggers predictor training when a new target model is deployed. Instead of running all predictors on control servers, which can create a hotspot, \sys{} colocates predictors with the worker servers that host the models they schedule. This makes prediction capacity scale with the cluster and gives predictors low-latency access to fresh prompt and runtime state.

Predictors are lightweight relative to the agentic inference tasks they support. For example, a 35M-parameter LLM predictor takes about 30\,ms on CPU and roughly 4\,ms on GPU, while target LLMs with hundreds of billions of parameters often take seconds to minutes. Most predictors therefore run on CPUs. Larger predictors can run on GPUs with modest memory overhead because they are orders of magnitude smaller than the target models.

Predictor weights are distributed from control servers under eventual consistency. Temporary inconsistency may cause short-lived accuracy drift, but prediction quality stabilizes as updates propagate. For reliability, predictor weights and associated context are periodically checkpointed. Sensitive prompt fields are protected through verified redaction before being stored or reused for adaptation.

\mypar{Handling high prediction traffic.}
Routers and scalers can generate high prediction traffic. Router-side prediction scales naturally with existing service replication. Scaler-side prediction is more challenging because one scaler may cover many model services. To reduce this load, \sys{} delegates heavy prompt parsing to upstream routers and lets the scaler use only the resulting prompt-aware representation with a lightweight MLP. This preserves prompt awareness while keeping scaler-side prediction overhead low.

\mypar{Failure handling.}
\sys{} preserves the failure model of the underlying infrastructure. Predictors are colocated with existing worker and control components rather than introduced as a separate critical service. On failure, predictor state is restored from checkpointed weights and context. If a predictor is temporarily unavailable, \sys{} falls back to the underlying scheduler policy, ensuring that prediction failures do not compromise service availability.

\section{Evaluation}
\label{sec:evaluation}

Our evaluation is organized around two complementary settings. First, controlled testbed experiments isolate \sys{}'s design choices and measure how much each scheduling mechanism contributes. Second, large-scale production deployments evaluate whether these gains hold under real workloads, heterogeneous resources, and operational constraints. We answer four questions.

\begin{itemize}[leftmargin=*]
\setlength{\itemsep}{1pt}
\setlength{\parskip}{0pt}

\item \textbf{Do individual \sys{} components improve scheduling decisions in a controlled testbed?}
Section~\ref{sec:microbenchmark} isolates each scheduler component. \sys{}'s distribution-aware router reduces P95 latency by up to \textbf{18\%} on \textit{Text-to-Video} and \textbf{28.3\%} on \textit{Deep Research} over Ray. Its distribution-aware scaler reduces latency by \textbf{7.8--13.7\%} on \textit{Text-to-Video} and by nearly \textbf{50\%} on \textit{Deep Research} over static provisioning.

\item \textbf{Do coordinated routers and scalers improve end-to-end performance?}
Section~\ref{sec:exp-e2e} evaluates \sys{} as a full system on the controlled testbed. It shows that request-time routing and window-level scaling are complementary: together, they reduce end-to-end P95 latency by up to \textbf{61.5\%} on structured pipelines and by \textbf{11--32\%} on open-ended agentic workloads, including OpenClaw and Coding Agent.

\item \textbf{Does \sys{} remain effective at production scale?}
Section~\ref{sec:exp-production} reports live deployments on production clusters with nearly one thousand heterogeneous GPUs and over one million CPU cores. \sys{} delivers up to \textbf{2$\times$} higher sustainable throughput under a fixed SLO and reduces P99 latency by \textbf{44--52\%} over production schedulers.

\item \textbf{What are \sys{}'s overhead, cost, and robustness?}
Section~\ref{sec:exp-ood} evaluates predictor overhead, resource cost, and robustness to drift. It shows that predictor overhead is negligible relative to the seconds-to-minutes execution time of served models, and that \sys{} detects a severe workload shift and recovers online within \textbf{100\,s}.
\end{itemize}

\subsection{Experiment Setup}


\begin{table*}[!th]
\centering
\footnotesize
\renewcommand{\arraystretch}{1.15}
\resizebox{\linewidth}{!}{
\setlength{\tabcolsep}{4pt}
\begin{tabular}{@{}>{\raggedright\arraybackslash}p{0.18\textwidth}
                >{\raggedright\arraybackslash}p{0.16\textwidth}
                >{\raggedright\arraybackslash}p{0.48\textwidth}
                >{\raggedright\arraybackslash}p{0.14\textwidth}@{}}
\toprule
\textbf{Category} & \textbf{Service} & \textbf{Pipeline} & \textbf{Input} \\
\midrule
\multirow{2}{=}{Structured LLM Pipelines}
& Deep Research
& Qwen3-32B (Plan/Summary); Two Qwen3-8B (Query)
& DR Bench~\cite{du2025deepresearch} \\

& Text-to-Video
& Qwen3-8B $\to$ Wan2.1-T2V-1.3B
& OpenVid-1M~\cite{nan2024openvid} \\

\addlinespace
\multirow{2}{=}{Open-ended Agentic Applications}
& OpenClaw
& \textbf{Dual-model setup:} Qwen3-Next-80B-A3B $+$ Qwen3-8B-VL; \textbf{Single-model setup:} Qwen3-Next-80B-A3B
& MCP Atlas~\cite{mcp-atlas} \\

& Coding Agent
& \textbf{Dual-model setup:} Qwen3-Next-80B-A3B (Plan) Qwen3-8B (Act); \textbf{Single-model setup:} Qwen3-Next-80B-A3B
& SWE-bench Pro~\cite{swebench-pro} (trace~\cite{dataclaw_peteromallet_2026}) \\

\addlinespace
\multirow{3}{=}{Production Deployments}
& Video OCR
& Internal Model (Detect); Internal Model (Recognize); Internal Model (Match)
& Prod.\ traffic \\

& Video Transcode
& Internal Model 
& Prod.\ traffic \\

& Entity Semantic Analysis
& Two Qwen3VL-8B (Recognization); Two Qwen3-omni-30B (Detection)
& Prod.\ traffic \\

\bottomrule
\end{tabular}
}
\caption{Evaluated workloads grouped by category.}
\label{tab:workloads}
\end{table*}

\mypar{Baselines.}
We compare \sys{} against three request-routing baselines, one calibration baseline in the controlled testbed, and one deployment baseline in production.

\begin{itemize}[leftmargin=*,itemsep=1pt,topsep=0pt,parsep=0pt,partopsep=0pt]

\item \textbf{Murakkab}~\cite{murakkab} is the latest agentic workload scheduling baseline (to appear in OSDI 26), in which it optimizes end-to-end latency for compound AI systems. Its optimizer relies on average latency estimates, representing global schedulers that use point estimates while discarding predictive uncertainty.

\item \textbf{Ray Core}~\cite{ray} is our \emph{production-default} baseline. It is widely used for complex AI workloads, and its round-robin dispatcher is a common default in our deployments under dynamic latency variation.

\item \textbf{Power-of-Two-Choices (PO2)}~\cite{mitzenmacher2002power} is our \emph{robust-heuristic} baseline. PO2 is low-overhead and competitive when accurate request-level prediction is unavailable, making it a useful reference for evaluating whether \sys{}'s prompt-aware predictions improve routing decisions.

\item \textbf{Random} is a \emph{calibration baseline}. It provides a lower-bound reference that separates the gains of informed scheduling from the gains of having any dispatch policy.

\item \textbf{Production scheduler} is our \emph{deployment baseline}. Production experiments compare against the scheduler already serving each workload. This is a strong baseline: it has been heavily optimized to use device and runtime signals for scheduling decisions. However, it is not prompt-aware and reduces scheduling state to point estimates, making it unable to exploit distribution-aware scheduling.
\end{itemize}


\mypar{Workloads.} As shown in Table~\ref{tab:workloads}, we evaluate \sys{} on seven services across three categories of increasing scheduling difficulty, covering diverse execution patterns, input sources, and both single- and dual-model modes. For structured LLM pipelines, Deep Research uses 20{,}000 samples from DR Bench\cite{du2025deepresearch} and Text-to-Video uses 2{,}000 samples from OpenVid-1M\cite{nan2024openvid}; for open-ended agentic applications, OpenClaw runs on the full MCP Atlas dataset\cite{mcp-atlas}, and Coding Agent uses the dataclaw-peteromallet\cite{dataclaw_peteromallet_2026} trace of SWE-bench Pro\cite{swebench-pro}. Production traffic include about \textbf{11 Millions (!)} records captured from real-world serving. The same predictors and scheduling mechanisms extend \sys{} to other services using the same served models.

\mypar{Testbed and implementation.}
Controlled experiments run on a GPU testbed with 128 NVIDIA H20 GPUs. Production experiments run on clusters with 320 NVIDIA H20 GPUs and 560 NVIDIA L20 GPUs, and production CPU clusters with over one million (!) cores.

Models are served with vLLM 0.16.0, and \sys{} runs as a scheduler plug-in over Ray, as described in Section~\ref{sec:implementation}. Each experiment reports its own deployment scale. Each \sys{} predictor pairs a parameter-reduced isomorphic semantic model with a lightweight MLP, trained from service traces using the procedure in Section~\ref{sec:memory}. Production results are collected from live deployments.

\mypar{Metrics.}
Our primary metric is request latency. Single-component microbenchmarks (Section~\ref{sec:microbenchmark}) report per-request latency for isolated scheduler components, while end-to-end and production experiments report full workflow latency. Because agentic workloads often run at low QPS with long and highly variable requests, a single poor scheduling decision can dominate the tail. We therefore report P95 and P99 latency alongside median latency. Where relevant, we report sustainable throughput under an SLO.

\subsection{\sys{}'s Impact on Scheduler Components}
\label{sec:microbenchmark}

We first isolate the two core scheduler components in \sys{}: the router and the scaler. These microbenchmarks run on the 128-GPU H20 cluster. When one component is evaluated, the other uses its default policy---round-robin routing or static provisioning with offline-profiled replica counts---so that the measured gain can be attributed to the component under test.

\begin{figure}[!t]
    \centering
    \includegraphics[width=0.95\linewidth]{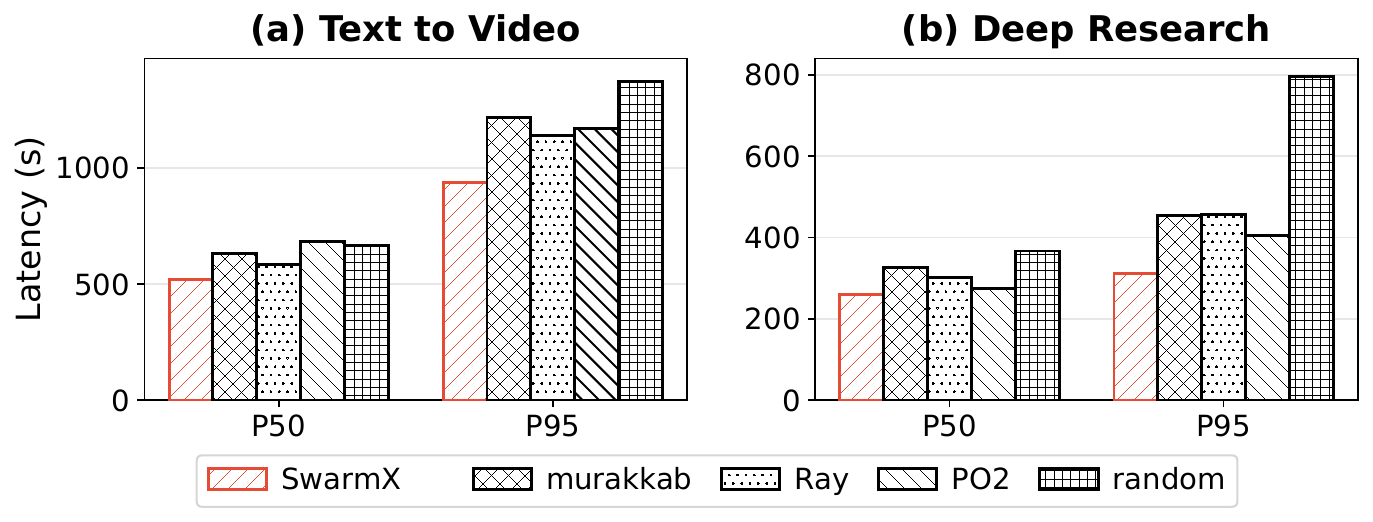}
    \caption{Router-only microbenchmark: P50 and P95 latency for (a) \textit{Text-to-Video} and (b) \textit{Deep Research}.}
    \Description{Two-panel bar chart of P50 and P95 latency comparing \sys{}'s router with round-robin and the other routing baselines on the Text-to-Video and Deep Research workloads.}
    \label{fig:router-agent-exp}
\end{figure}

\mypar{Router: distribution-aware dispatch.}
We isolate the router while holding instance allocation fixed at the offline-profiled replica counts. \autoref{fig:router-agent-exp}(a) reports P50 and P95 workflow latency for \textit{Text-to-Video}. \sys{} reduces P50/P95 latency by \textbf{11.0\%/18.0\%} over Ray. Murakkab, although also prediction-based, performs worse than Ray by \textbf{18.0\%/23.1\%}. Its point-estimate latency model cannot represent the broad, multi-modal latency distribution caused by variable diffusion-iteration counts. As a result, it cannot reliably distinguish a queue with many short requests from one with a few long requests, leading to costly misrouting. \sys{} avoids this problem by preserving prediction distributions for more effective routing.

For \textit{Deep Research} (\autoref{fig:router-agent-exp}(b)), \sys{} reduces P50/P95 latency by \textbf{29.0\%/28.3\%} over Ray and by \textbf{20.3\%/31.2\%} over Murakkab. The gain is larger than in \textit{Text-to-Video} because \textit{Deep Research} is dominated by \texttt{thinking} and \texttt{summarize} calls whose latency is highly prompt-dependent. These semantic-dependent latency differences are captured by \sys{}'s semantic model. This matches our design rationale: the wider the per-request latency spread, the more a distribution-aware router benefits over policies that ignore predictive uncertainty.

\begin{figure}[!t]
    \centering
    \includegraphics[width=0.95\linewidth]{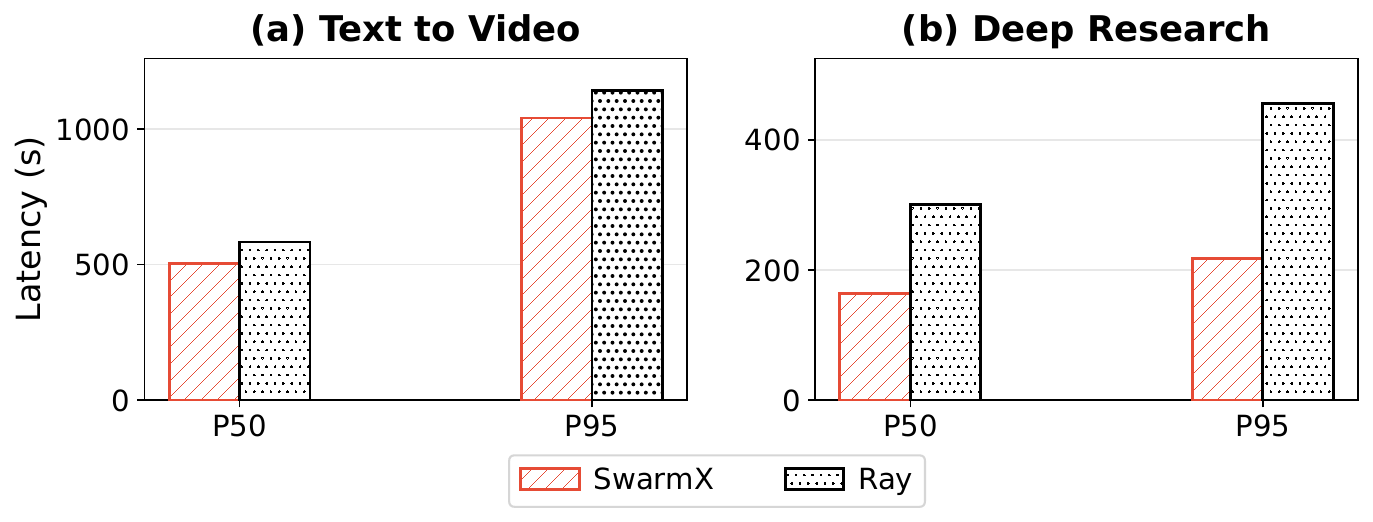}
    \caption{Scaler-only microbenchmark: P50 and P95 latency for (a) \textit{Text-to-Video} and (b) \textit{Deep Research}.}
    \Description{Two-panel bar chart of P50 and P95 latency comparing \sys{}'s scaler with static provisioning on the Text-to-Video and Deep Research workloads.}
    \label{fig:runtime-scaler-exp}
\end{figure}

\mypar{Scaler: structure-aware provisioning.}
We next isolate the scaler while routing requests round-robin, so that the measured gain comes only from provisioning decisions. The baseline is static provisioning, with replica counts fixed from offline profiling. For \textit{Text-to-Video} (\autoref{fig:runtime-scaler-exp}(a)), \sys{} reduces latency by \textbf{7.8--13.7\%} over static provisioning. Variable diffusion-iteration depth shifts aggregate demand over time, and proactive scaling tracks these changes better than a fixed allocation.

For \textit{Deep Research} (\autoref{fig:runtime-scaler-exp}(b)), the gain rises to approximately \textbf{50\%} across percentiles. \textit{Deep Research} has stronger structure dynamics: both fan-out degree and call depth vary with prompt semantics. Static provisioning is therefore frequently misaligned with actual demand. \sys{} forecasts demand from predicted call structure and provisions ahead of queue buildup, rather than reacting only after queues have formed.

\mypar{Ablation study.}
We also ablated two design choices: the semantic model and distribution-aware scheduling. Due to space constraints, we omit the detailed results. The main findings are that the semantic model is critical for capturing prompt semantics, which the MLP alone cannot reliably infer, and that distribution-aware scheduling provides better tail-latency control than point-estimate-based scheduling.

\subsection{\sys{}'s End-to-End Performance}
\label{sec:exp-e2e}

\begin{figure}[!t]
    \centering
    \includegraphics[width=0.95\linewidth]{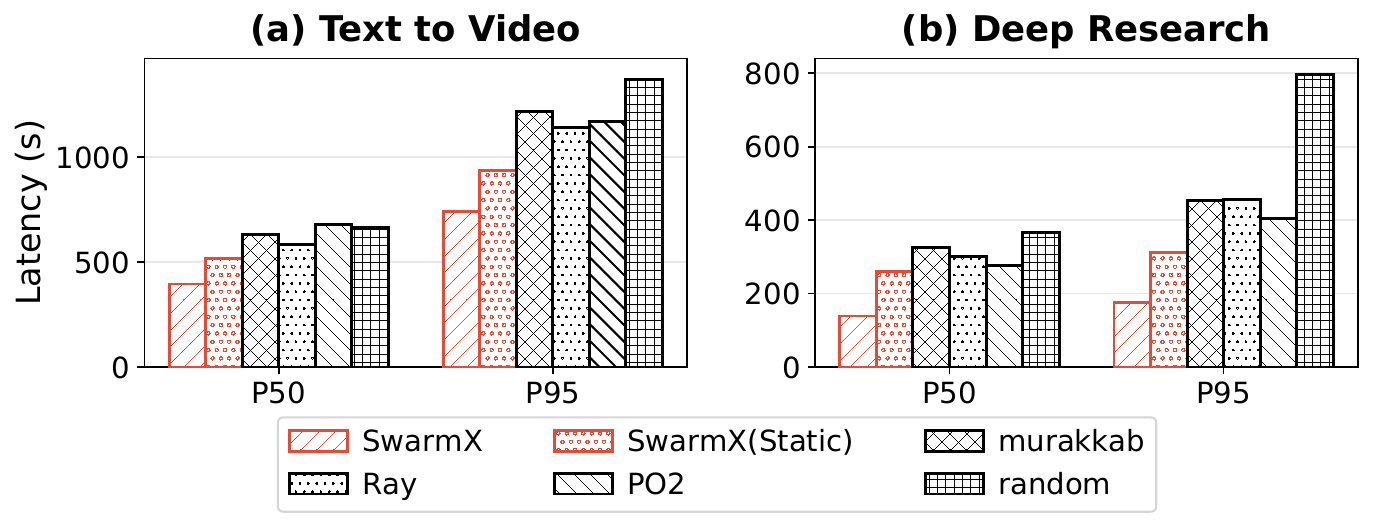}
    \caption{End-to-end P50 and P95 latency for (a) \textit{Text-to-Video} and (b) \textit{Deep Research}. \sys{}\,(Static) runs the \sys{} router with the scaler disabled.}
    \Description{Two-panel bar chart of P50 and P95 end-to-end latency comparing \sys{}, \sys{}\,(Static), Ray, PO2, Murakkab, and Random on the Text-to-Video and Deep Research workloads.}
    \label{fig:makespan_cdf}
\end{figure}

\begin{figure}[!t]
    \centering
    \includegraphics[width=0.95\linewidth]{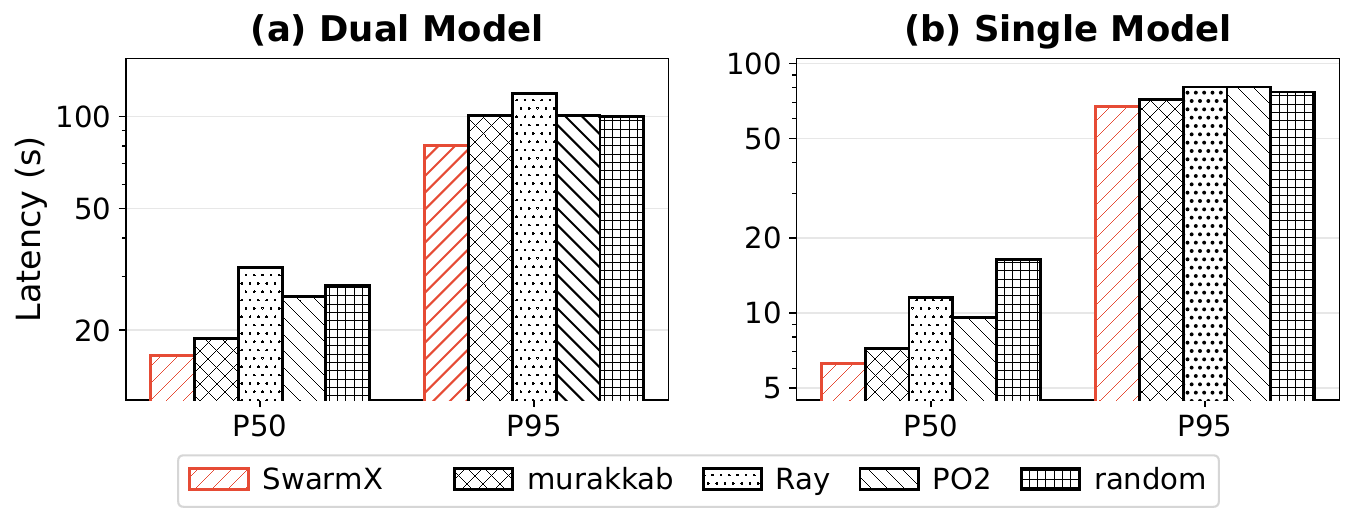}
    \caption{End-to-end P50 and P95 latency of OpenClaw for (a) dual-model setup and (b) single-model setup.}
    \Description{Two-panel bar chart of P50 and P95 end-to-end latency of OpenClaw under dual-model and single-model setup, comparing \sys{} with Murakkab, Ray, PO2, and Random.}
    \label{fig:openclaw_cdf}
\end{figure}

\begin{figure}[!t]
    \centering
    \includegraphics[width=0.95\linewidth]{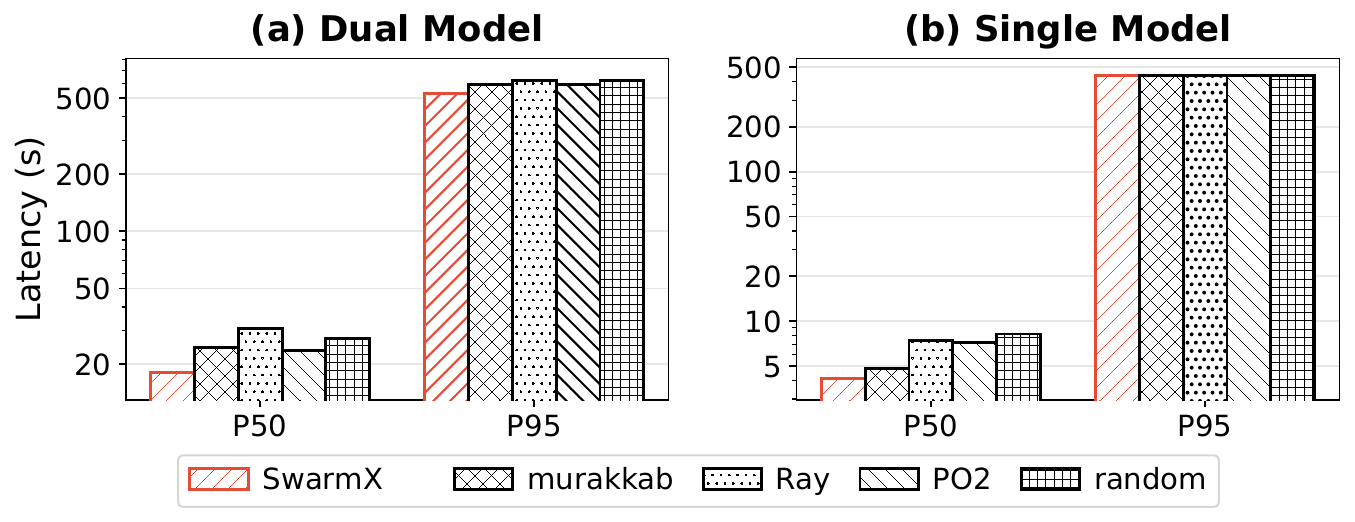}
    \caption{End-to-end P50 and P95 latency of Coding Agent for (a) dual-model setup and (b) single-model setup.}
    \Description{Two-panel bar chart of P50 and P95 end-to-end latency of Coding Agent under dual-model and single-model setup, comparing \sys{} with Murakkab, Ray, PO2, and Random.}
    \label{fig:claude_code_cdf}
\end{figure}

We now evaluate the full system, with the router and scaler running together through the coordinated control path of Section~\ref{sec:interface}.
\autoref{fig:makespan_cdf} reports end-to-end latency for the two structured pipelines, \textit{Text-to-Video} and \textit{Deep Research}, against all baselines.
It also includes \sys{}\,(Static)---the \sys{} router with the scaler disabled and replicas statically provisioned---so that the figure isolates the scaler's contribution to end-to-end latency.

\sys{} achieves the lowest P50 and P95 latency on both pipelines.
Against Ray Core, it reduces P50/P95 latency by \textbf{32.1\%/34.9\%} for \textit{Text-to-Video} and by \textbf{53.7\%/61.5\%} for \textit{Deep Research}.
The gain is larger for \textit{Deep Research} because its pipeline varies in both fan-out and iteration depth, giving joint control more to correct, whereas \textit{Text-to-Video}'s single-stage diffusion exposes a narrower range of structure dynamics.

Coordinating both components, rather than running either alone, is what produces these gains.
The gap between \sys{} and \sys{}\,(Static) in \autoref{fig:makespan_cdf} isolates the scaler's effect: enabling the scaler on top of the \sys{} router reduces \textit{Text-to-Video} P95 latency by a further \textbf{20.7\%}, and reduces \textit{Deep Research} P95 latency by a further \textbf{43.8\%}.
This shows that being request-time and structure-aware are complementary: each corrects a different source of latency, and neither alone matches the full system.

\textit{OpenClaw} and \textit{Coding Agent} test whether this result holds beyond structured pipelines.
Both are open-ended agentic workloads: model roles, latency distributions, and call structures are decided at runtime rather than fixed by a stage graph, which makes parallelism harder to predict and scheduling decisions more challenging compared to the earlier settings.
\autoref{fig:openclaw_cdf} and \autoref{fig:claude_code_cdf} compare \sys{} against the same baselines in the dual-model and single-model deployments of Table~\ref{tab:workloads}.

\sys{} retains strong gains. In the dual-model setup, it reduces P50/P95 latency by \textbf{25.9\%/11.0\%} for \textit{Coding Agent} and by \textbf{11.9\%/19.8\%} for \textit{OpenClaw} compared to Murakkab, and by \textbf{40.8\%/14.9\%} and \textbf{48.6\%/32.4\%} compared to Ray, respectively. In the single-model setup, it reduces P50 latency by \textbf{13.8\%} for \textit{Coding Agent} and \textbf{12.8\%} for \textit{OpenClaw} compared to Murakkab, and by \textbf{44.49\%} and \textbf{45.5\%} compared to Ray. For P95 latency, \sys{} reduces latency on \textit{OpenClaw} by \textbf{5.8\%} compared to Murakkab and by \textbf{16.2\%} compared to Ray, while maintaining similar performance on \textit{Coding Agent}, whose workload distribution is more homogeneous than that of \textit{OpenClaw}.

That \sys{} remains effective even when the call structure is determined entirely at runtime indicates that its predictors generalize beyond well-structured services.

\subsection{Large-Scale Production Deployment}
\label{sec:exp-production}

\begin{figure}[!t]
    \centering

    \begin{minipage}[t]{0.47\linewidth}
        \centering
        \includegraphics[
            width=\linewidth,
            height=0.28\textheight,
            keepaspectratio
        ]{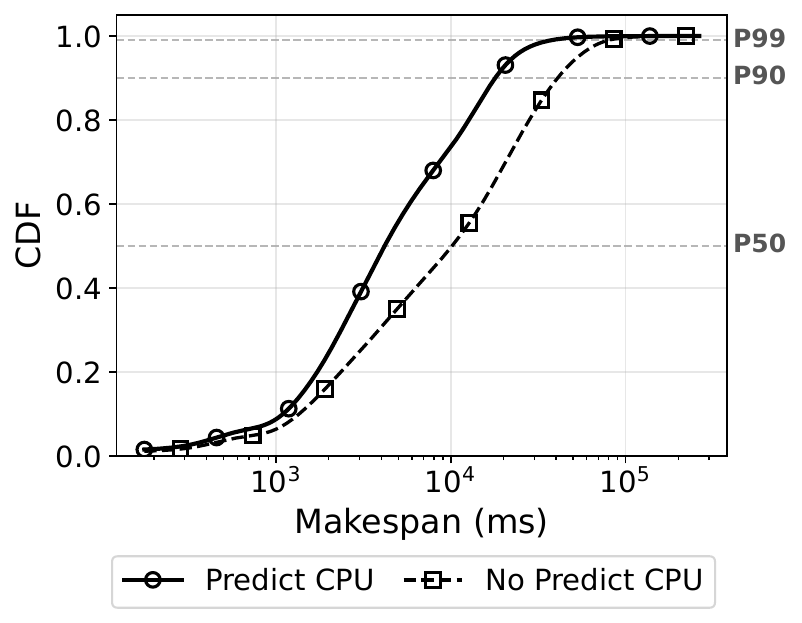}
        \captionof{figure}{End-to-end latency CDF for the \textit{Video OCR} service on a CPU cluster.}
        \Description{Latency CDF on a log scale comparing \sys{} with the production scheduler for the Video OCR service running on a large CPU cluster.}
        \label{fig:ocr_makespan}
    \end{minipage}
    \hfill
    \begin{minipage}[t]{0.47\linewidth}
        \centering
        \includegraphics[
            width=\linewidth,
            height=0.28\textheight,
            keepaspectratio
        ]{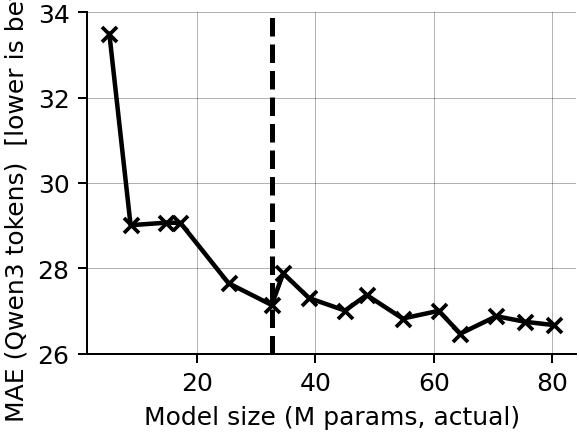}
        \captionof{figure}{Select semantic model based on accuracy--size tradeoff.
        }
        \Description{Line chart showing prediction accuracy versus parameter count for Qwen3-family semantic models, with accuracy improving and saturating as size grows.}
        \label{fig:semantic_model_size_sweep}
    \end{minipage}

\end{figure}

We next evaluate whether \sys{} remains effective in production, where clusters are larger, more heterogeneous, and more volatile than our controlled testbed. We have deployed \sys{} on three high-traffic internal services---\textit{Video OCR}, \textit{Entity Semantic Analysis}, and \textit{Video Transcode}---spanning CPU clusters with over one million cores and heterogeneous GPU clusters with nearly one thousand devices. Across these deployments, \sys{} reduces tail latency by up to \textbf{52\%} or increases sustainable throughput by up to \textbf{2$\times$} over the production scheduler.

\mypar{Multiple models on large CPU clusters.}
\textit{Video OCR} runs a three-stage detect--recognize--match pipeline on a CPU cluster with tens of thousands of cores. In this fragmented, high-volume environment, \sys{} reduces P50 latency by \textbf{59.6\%} and P99 latency by \textbf{48.38\%} relative to the production scheduler (\autoref{fig:ocr_makespan}). This result shows that \sys{} is not limited to GPU-centric serving: prediction-driven scheduling also benefits CPU services with input-dependent latency variation and multi-stage execution.

\mypar{Multiple models on heterogeneous GPU clusters.}
\textit{Entity Semantic Analysis} runs on a heterogeneous GPU cluster with 320 NVIDIA H20 GPUs and 560 NVIDIA L20 GPUs. The scheduler must balance hardware capability, request complexity, runtime load, and hardware priority. In a \emph{capacity test}---raising load until the SLO is first violated---\sys{} sustains approximately \textbf{2$\times$} the throughput of the prior production scheduler under the same SLO. It achieves this without hand-written hardware rules: GPU type is treated as a device feature, and the predictor learns the H20/L20 performance gap from traces.

\begin{figure}[!t]
    \centering
    \includegraphics[width=\linewidth]{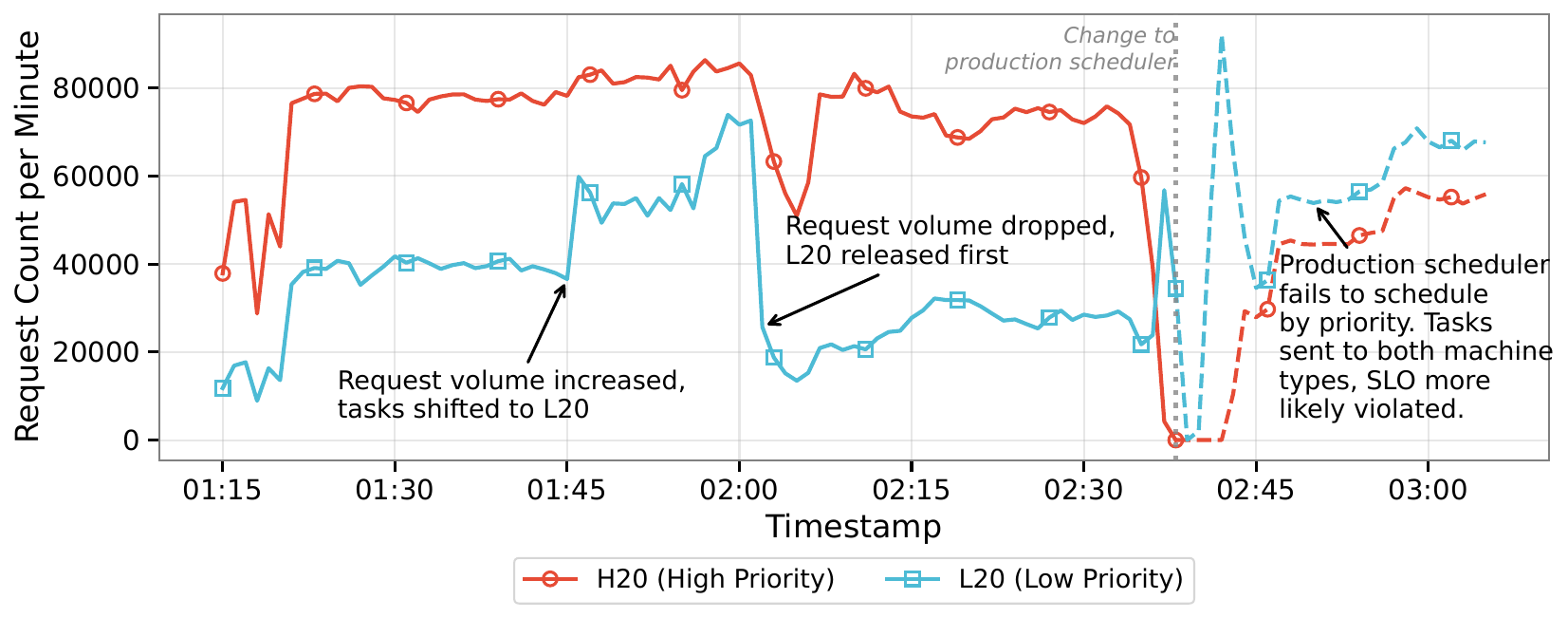}
    \caption{Priority-aware routing on a heterogeneous cluster. \sys{} keeps load on the high-priority H20 pool and spills to L20 only under high volume; after the switch to the production scheduler (dashed line), work is sent to both pools regardless of priority.}
    \Description{Bar/area chart showing the distribution of dispatched requests across H20 and L20 GPU pools over time, with H20 preferred until saturation pushes load to L20.}
    \label{fig:video_cleaning_gpu}
\end{figure}

\mypar{Priority-aware routing on heterogeneous GPUs.}
We further examine \textit{Entity Recognition}, one stage of \textit{Entity Semantic Analysis}, to understand how \sys{} handles hardware preferences. \autoref{fig:video_cleaning_gpu} traces request placement over a production window. \sys{} keeps work on the higher-priority H20 pool and spills to the lower-priority L20 pool only when H20 saturates; as load falls, it drains L20 first. After switching back to the production scheduler, shown to the right of the dashed line, requests are spread across both pools regardless of priority, and SLO violations increase.

In a separate fixed-load comparison on the same cluster at production traffic volume, \sys{} reduces P99 latency by about \textbf{50\%} and increases throughput by about \textbf{40\%} under the same SLO. This case study shows that \sys{} can incorporate deployment-specific device and runtime preferences through its feature and action interfaces, while preserving the same predictor-driven scheduling framework.

\mypar{CPU-intensive workloads: video transcode and tool calls.}
\sys{}'s gains are not limited to GPU-intensive workloads. \textit{Video Transcode} is a CPU-only production service running on over one million cores. It is not AI-native and exposes no multi-model workflow graph, yet its per-request latency still varies strongly with input. On production traffic, \sys{} reduces P99 latency by \textbf{44--52\%}, showing that prediction-driven scheduling applies to services with input-dependent latency variation beyond agentic pipelines. We also apply \sys{} to CPU-intensive tool-call services used inside agentic workloads and observe consistent gains, but have to omit them due to page limit.



\subsection{System Overhead, Cost, and Robustness}
\label{sec:exp-ood}

The gains reported so far are only useful if \sys{} is cheap to run and stays accurate as conditions change.
This subsection answers two questions: is prediction cheap enough to sit on the online scheduling path, and can \sys{} recover when the workload drifts away from what its predictors were trained on?


\mypar{Selecting a semantic model.}
Within the neural predictor, the semantic model, dominates predictor cost while MLP stays lightweight. Therefore, the semantic model's size sets the overhead budget: it must stay small enough for the scheduling path while still capturing prompt semantics.
We size it with a sweep over the Qwen3 model family.
We instantiate several parameter-reduced isomorphic variants that preserve the Qwen3 architectural shape, replace the final layer with an output-length prediction head, train each with the same pipeline as the serving model family (Section~\ref{sec:memory}), and measure output-length prediction error.
\autoref{fig:semantic_model_size_sweep} shows that error drops sharply with size and then saturates, with the 35M model already achieving a low error of 27.89, \sys{} therefore selects the smallest variant past the knee, preserving accuracy while avoiding unnecessary training and inference overhead.


\begin{table}[!t]
    \centering
    \scriptsize
    \renewcommand{\arraystretch}{1.15}
    \begin{tabular}{@{}p{0.28\columnwidth}
                    p{0.30\columnwidth}
                    p{0.32\columnwidth}@{}}
        \toprule
        \textbf{} & \textbf{Wan2.1-T2V-1.3B} & \textbf{Qwen3-8B} \\
        \midrule

        \textbf{Served-model latency}
        & 17--137\,s
        & 0.7--100+\,s \\

        \textbf{Prediction time}
        & $<$1\,ms
        & 30\,ms (CPU), $\sim$4\,ms (GPU) \\

        \textbf{Memory}
        & 261\,KB (66K parameters)
        & $\sim$100\,MB (35M parameters) \\

        \bottomrule
    \end{tabular}
    \caption{\sys{}'s predictor overhead and footprint for two representative target models.}
    \label{tab:agent-overhead-and-cost}
    \vspace{-.1in}
\end{table}


\mypar{Overhead and cost.}
\autoref{tab:agent-overhead-and-cost} shows that \sys{}'s predictors are orders of magnitude cheaper than the models they schedule.
For the Wan2.1-T2V-1.3B diffusion model, a \textbf{66K}-parameter predictor suffices: it runs in under \textbf{1\,ms} and occupies \textbf{261\,KB}.
A language model's output is more challenging to predict than a diffusion model's---it depends on prompt semantics beyond frame count and resolution---so for Qwen3-8B \sys{} uses a larger 35M-parameter predictor, which runs in about \textbf{30\,ms} on CPU, roughly \textbf{4\,ms} on GPU, and occupies under 100\,MB.
In both cases prediction takes milliseconds while the target model takes seconds to minutes. As a result, the predictor can be colocated with execution without becoming the bottleneck.



\begin{figure}[!t]
    \centering
    \includegraphics[width=0.65\linewidth]{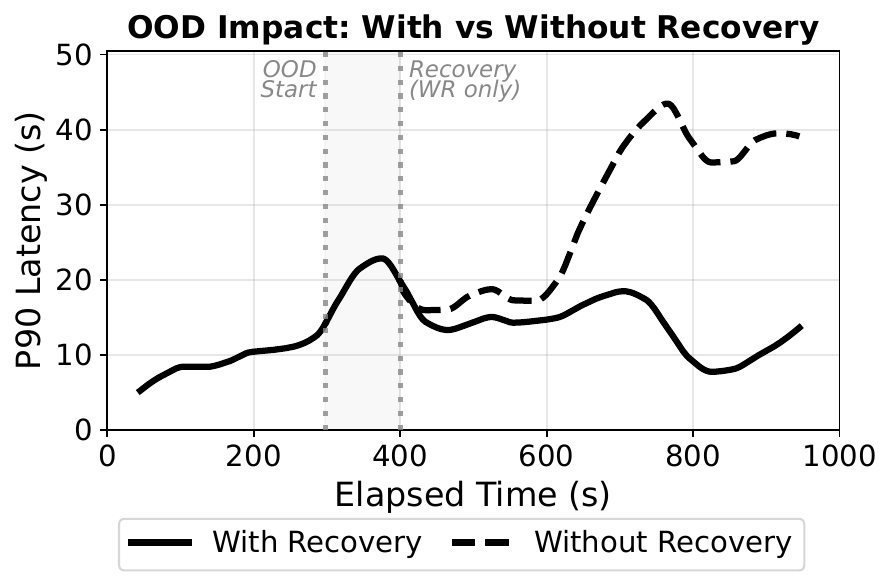}
    \caption{P90 latency through a 71\% resource capacity loss. With OOD-triggered retraining, \sys{} recovers to near pre-shift latency; without it, tail latency keeps rising.}
    \Description{Time-series plot of P90 latency comparing \sys{} with and without online retraining after a simulated out-of-distribution capacity loss; the with-recovery line returns toward baseline while the without-recovery line keeps rising.}
    \label{fig:ood_on_router}
    \vspace{-.1in}
\end{figure}

\mypar{Robustness under severe drift.}
\sys{} must also stay accurate when the runtime environment changes.
We test this on \textit{Deep Research} with Qwen3-32B on an 80-GPU cluster, injecting a severe shift by cutting each GPU's available resources for a \textbf{71\%} aggregate capacity loss.
\autoref{fig:ood_on_router} compares \sys{} with and without online adaptation.
Without retraining, the stale predictor misroutes requests and P90 latency climbs to more than \textbf{40\,s}.
With OOD-triggered adaptation, \sys{} detects the shift, retrains the distribution predictor, and holds P90 latency below \textbf{20\,s} (Section~\ref{sec:memory}); detection and retraining complete within \textbf{100\,s}, showing that \sys{} can recover autonomously from severe environmental shift.


\section{Production and Operation Experience}
\label{sec:experience}

\sys{} has been deployed and evolved in production for two years across large CPU clusters, heterogeneous GPU clusters, and hundreds of AI applications. We summarize the main lessons from operating it at scale.

\mypar{Compatibility enables adoption; the agent framework sustains it.}
A production scheduler is difficult to adopt if application teams must rewrite serving code. \sys{} therefore preserves existing Ray and Kubernetes programming interfaces and runs as a plug-in. Compatibility alone, however, is not sufficient. Teams also value \sys{} for packaging neural predictors and their management mechanisms into a modular agent framework with clean interfaces to the underlying cluster infrastructure. This design allows infrastructure teams to embed neural predictors into existing systems with minimal modification. The framework further decomposes key operations, such as prediction, monitoring, and retraining, into asynchronous tasks, allowing \sys{} to run efficiently on existing cluster engines such as Ray.

\mypar{Schedule at role granularity, coordinate at application scope.}
Our initial deployment scaled applications by launching multiple Ray clusters. This improved isolation and horizontal scalability, but resource allocation remained coarse-grained, operating only at the cluster or actor level. As applications became more agentic, their stages began to exhibit distinct resource demands, elasticity patterns, and bottlenecks. \sys{} therefore represents each application as a set of roles, where each role has its own resource demand, scheduling policy, and elasticity constraints. This representation allows prefill, decode, and CPU-intensive tool execution to be managed independently.

At the same time, optimizing a single role may not reduce end-to-end application latency if it only shifts pressure to another role. \sys{} therefore treats the application as the coordination boundary: an application-level controller coordinates resources across roles, where each role may correspond to a model or a model stage such as prefill or decode; local routers and scalers make role-level decisions; and \sys{} maintains an aggregate view of latency, resource usage, and cost across the full application.

\mypar{Heterogeneity support is critical for deployment.}
Agentic AI workloads combine GPU inference, CPU-side processing, sandboxed tool execution, and external service calls. \sys{} avoids hand-written scheduling logic for each CPU, GPU, or GPU type. Instead, it represents hardware and runtime properties as predictor features, allowing the data-driven predictors to adapt across diverse hardware configurations and support unified CPU--GPU scheduling. Application teams particularly valued this capability because it enabled decisions that were previously difficult to realize, such as using heterogeneous GPUs within a large workflow or avoiding over-scaling GPU replicas when CPU-side work was the actual bottleneck.


\section{Related Work}
\label{sec:related}

\mypar{Agent workflow frameworks.}
LangGraph~\cite{langgraph}, AutoGen~\cite{autogen}, and CrewAI~\cite{crewai} provide programming abstractions for chaining LLM calls, multi-agent collaboration, and tool use. They define \emph{what} an agentic workflow looks like, but leave \emph{where} and \emph{when} each call runs to the underlying infrastructure. \sys{} sits beneath these frameworks as a layer that performs runtime routing and scaling.

\mypar{LLM inference optimization.}
DistServe~\cite{distserve}, Llumnix~\cite{llumnix}, Orca~\cite{orca}, and vLLM~\cite{vllm} optimize inference within a model replica through techniques such as prefill/decode disaggregation, iteration-level batching, and paged KV-cache management. \sys{} operates one level above these systems: it treats optimized replicas as scheduling targets and coordinates routing and scaling across multi-model agentic workflows. The two directions are complementary.

\mypar{Cluster and predictive scheduling.}
Ray~\cite{ray}, Kubernetes~\cite{kubernetes}, and related cluster managers provide robust deployment and scheduling interfaces for distributed services. Predictive schedulers such as PSC~\cite{faisal2024will} and Cilantro~\cite{288542} show that learned or statistical models can improve scheduling. \sys{} builds on this direction but targets agentic AI serving, where scheduling depends on prompt semantics, target-model behavior, device heterogeneity, and runtime state. It also exposes distributional predictions to routers and scalers rather than reducing predictions to point estimates.

\mypar{Scheduling for compound and agentic AI.}
Recent systems such as ORION~\cite{orion}, Pie~\cite{gim2025pie}, Parrot~\cite{lin2024parrot}, Murakkab~\cite{murakkab}, and Pythia~\cite{yu2026pythiaexploitingworkflowpredictability} study scheduling for compound or agentic AI workloads. \sys{} differs by jointly addressing prompt-aware prediction, distribution-aware routing and scaling, online predictor adaptation, and infrastructure integration through a scheduler-agent framework.

\section{Conclusion}
\label{sec:conclusion}

This work establishes agentic scheduling as a practical foundation for serving large-scale agentic applications. \sys{} formulates routing and scaling as neural-prediction-driven scheduling problems, uses scheduling-specific predictors to improve router and scaler decisions, and manages these predictors through a scheduler-agent framework that integrates cleanly with existing scheduling and model-serving infrastructure. Our results show that \sys{} significantly improves latency across critical agentic AI applications where state-of-the-art schedulers fall short. More broadly, agentic scheduling opens a new direction for AI-driven cluster scheduling, and the soon-to-be open-sourced \sys{} lays the groundwork for a new generation of intelligent cluster schedulers.

\bibliographystyle{ACM-Reference-Format}
\bibliography{sample}

@inproceedings {distserve,
author = {Yinmin Zhong and Shengyu Liu and Junda Chen and Jianbo Hu and Yibo Zhu and Xuanzhe Liu and Xin Jin and Hao Zhang},
title = {{DistServe}: Disaggregating Prefill and Decoding for Goodput-optimized Large Language Model Serving},
booktitle = {18th USENIX Symposium on Operating Systems Design and Implementation (OSDI 24)},
year = {2024},
isbn = {978-1-939133-40-3},
address = {Santa Clara, CA},
pages = {193--210},
url = {https://www.usenix.org/conference/osdi24/presentation/zhong-yinmin},
publisher = {USENIX Association},
month = jul
}

@misc{yu2026pythiaexploitingworkflowpredictability,
      title={Pythia: Exploiting Workflow Predictability for Efficient Agent-Native LLM Serving},
      author={Shan Yu and Junyi Shu and Yuanjiang Ni and Kun Qian and Xue Li and Yang Wang and Jinyuan Zhang and Ziyi Xu and Shuo Yang and Lingjun Zhu and Ennan Zhai and Qingda Lu and Jiarong Xing and Youyou Lu and Xin Jin and Xuanzhe Liu and Harry Xu},
      year={2026},
      eprint={2604.25899},
      archivePrefix={arXiv},
      primaryClass={cs.MA},
      url={https://arxiv.org/abs/2604.25899},
}

@inproceedings {llumnix,
author = {Biao Sun and Ziming Huang and Hanyu Zhao and Wencong Xiao and Xinyi Zhang and Yong Li and Wei Lin},
title = {Llumnix: Dynamic Scheduling for Large Language Model Serving},
booktitle = {18th USENIX Symposium on Operating Systems Design and Implementation (OSDI 24)},
year = {2024},
isbn = {978-1-939133-40-3},
address = {Santa Clara, CA},
pages = {173--191},
url = {https://www.usenix.org/conference/osdi24/presentation/sun-biao},
publisher = {USENIX Association},
month = jul
}

@misc{langgraph,
  author = {LangChain},
  title = {LangGraph: Build resilient language agents as graphs},
  url ={https://github.com/langchain-ai/langgraph},
  year = {2024}
}

@misc{autogen,
  author = {Microsoft},
  title = {AutoGen: Enabling Next-Gen LLM Applications via Multi-Agent Conversation},
  url =  {https://github.com/microsoft/autogen},
  year = {2024}
}

@misc{crewai,
  author = {CrewAI},
  title = {CrewAI: Framework for orchestrating role-playing, autonomous AI agents},
  url = {https://github.com/joaomdmoura/crewAI},
  year = {2024}
}

@misc{murakkab,
      title={Murakkab: Resource-Efficient Agentic Workflow Orchestration in Cloud Platforms},
      author={Gohar Irfan Chaudhry and Esha Choukse and Haoran Qiu and Íñigo Goiri and Rodrigo Fonseca and Adam Belay and Ricardo Bianchini},
      year={2025},
      eprint={2508.18298},
      archivePrefix={arXiv},
      primaryClass={cs.MA},
      url={https://arxiv.org/abs/2508.18298},
}

@inproceedings {orca,
author = {Gyeong-In Yu and Joo Seong Jeong and Geon-Woo Kim and Soojeong Kim and Byung-Gon Chun},
title = {Orca: A Distributed Serving System for {Transformer-Based} Generative Models},
booktitle = {16th USENIX Symposium on Operating Systems Design and Implementation (OSDI 22)},
year = {2022},
isbn = {978-1-939133-28-1},
address = {Carlsbad, CA},
pages = {521--538},
url = {https://www.usenix.org/conference/osdi22/presentation/yu},
publisher = {USENIX Association},
month = jul
}

@inproceedings{vllm,
author = {Kwon, Woosuk and Li, Zhuohan and Zhuang, Siyuan and Sheng, Ying and Zheng, Lianmin and Yu, Cody Hao and Gonzalez, Joseph and Zhang, Hao and Stoica, Ion},
title = {Efficient Memory Management for Large Language Model Serving with PagedAttention},
year = {2023},
isbn = {9798400702297},
publisher = {Association for Computing Machinery},
address = {New York, NY, USA},
url = {https://doi.org/10.1145/3600006.3613165},
doi = {10.1145/3600006.3613165},
booktitle = {Proceedings of the 29th Symposium on Operating Systems Principles},
pages = {611–626},
numpages = {16},
location = {Koblenz, Germany},
series = {SOSP '23}
}

@misc{du2025deepresearch,
      title={DeepResearch Bench: A Comprehensive Benchmark for Deep Research Agents},
      author={Mingxuan Du and Benfeng Xu and Chiwei Zhu and Xiaorui Wang and Zhendong Mao},
      year={2025},
      eprint={2506.11763},
      archivePrefix={arXiv},
      primaryClass={cs.CL},
      url={https://arxiv.org/abs/2506.11763},
}

@misc{anthropic2025claudecode,
  author       = {{Anthropic}},
  title        = {Claude Code: Anthropic's agentic coding system},
  year         = {2025},
  url          = {https://www.anthropic.com/product/claude-code},
  note         = {Accessed: 2026-05-11}
}

@misc{openclaw,
  author       = {{OpenClaw Contributors}},
  title        = {OpenClaw: Open-Source Multi-Agent Framework},
  year         = {2025},
  url          = {https://github.com/open-claw/openclaw}
}

@article{nan2024openvid,
  title={OpenVid-1M: A Large-Scale High-Quality Dataset for Text-to-video Generation},
  author={Nan, Kepan and Xie, Rui and Zhou, Penghao and Fan, Tiehan and Yang, Zhenheng and Chen, Zhijie and Li, Xiang and Yang, Jian and Tai, Ying},
  journal={arXiv preprint arXiv:2407.02371},
  year={2024}
}

@misc{leviathan2025generativeui,
      title={Generative UI: LLMs are Effective UI Generators},
      author={Yaniv Leviathan and Dani Valevski and Matan Kalman and Danny Lumen and Eyal Segalis and Eyal Molad and Shlomi Pasternak and Vishnu Natchu and Valerie Nygaard and Srinivasan and Venkatachary and James Manyika and Yossi Matias},
      year={2026},
      eprint={2604.09577},
      archivePrefix={arXiv},
      primaryClass={cs.HC},
      url={https://arxiv.org/abs/2604.09577},
}

@inproceedings{
liu2025toolace,
title={Tool{ACE}: Winning the Points of {LLM} Function Calling},
author={Weiwen Liu and Xu Huang and Xingshan Zeng and xinlong hao and Shuai Yu and Dexun Li and Shuai Wang and Weinan Gan and Zhengying Liu and Yuanqing Yu and Zezhong WANG and Yuxian Wang and Wu Ning and Yutai Hou and Bin Wang and Chuhan Wu and Wang Xinzhi and Yong Liu and Yasheng Wang and Duyu Tang and Dandan Tu and Lifeng Shang and Xin Jiang and Ruiming Tang and Defu Lian and Qun Liu and Enhong Chen},
booktitle={The Thirteenth International Conference on Learning Representations},
year={2025},
url={https://openreview.net/forum?id=8EB8k6DdCU}
}

@article{wei2025deepseek,
  title={DeepSeek-OCR: Contexts Optical Compression},
  author={Wei, Haoran and Sun, Yaofeng and Li, Yukun},
  journal={arXiv preprint arXiv:2510.18234},
  year={2025}
}

@misc{cui2025paddleocr30technicalreport,
      title={PaddleOCR 3.0 Technical Report},
      author={Cheng Cui and Ting Sun and Manhui Lin and Tingquan Gao and Yubo Zhang and Jiaxuan Liu and Xueqing Wang and Zelun Zhang and Changda Zhou and Hongen Liu and Yue Zhang and Wenyu Lv and Kui Huang and Yichao Zhang and Jing Zhang and Jun Zhang and Yi Liu and Dianhai Yu and Yanjun Ma},
      year={2025},
      eprint={2507.05595},
      archivePrefix={arXiv},
      primaryClass={cs.CV},
      url={https://arxiv.org/abs/2507.05595},
}

@inproceedings {288542,
author = {Romil Bhardwaj and Kirthevasan Kandasamy and Asim Biswal and Wenshuo Guo and Benjamin Hindman and Joseph Gonzalez and Michael Jordan and Ion Stoica},
title = {Cilantro: {Performance-Aware} Resource Allocation for General Objectives via Online Feedback},
booktitle = {17th USENIX Symposium on Operating Systems Design and Implementation (OSDI 23)},
year = {2023},
isbn = {978-1-939133-34-2},
address = {Boston, MA},
pages = {623--643},
url = {https://www.usenix.org/conference/osdi23/presentation/bhardwaj},
publisher = {USENIX Association},
month = jul
}

@inproceedings{omega,
author = {Schwarzkopf, Malte and Konwinski, Andy and Abd-El-Malek, Michael and Wilkes, John},
title = {Omega: flexible, scalable schedulers for large compute clusters},
year = {2013},
isbn = {9781450319942},
publisher = {Association for Computing Machinery},
address = {New York, NY, USA},
url = {https://doi.org/10.1145/2465351.2465386},
doi = {10.1145/2465351.2465386},
booktitle = {Proceedings of the 8th ACM European Conference on Computer Systems},
pages = {351–364},
numpages = {14},
location = {Prague, Czech Republic},
series = {EuroSys '13}
}

@inproceedings {orion,
author = {Ashraf Mahgoub and Edgardo Barsallo Yi and Karthick Shankar and Sameh Elnikety and Somali Chaterji and Saurabh Bagchi},
title = {{ORION} and the Three Rights: Sizing, Bundling, and Prewarming for Serverless {DAGs}},
booktitle = {16th USENIX Symposium on Operating Systems Design and Implementation (OSDI 22)},
year = {2022},
isbn = {978-1-939133-28-1},
address = {Carlsbad, CA},
pages = {303--320},
url = {https://www.usenix.org/conference/osdi22/presentation/mahgoub},
publisher = {USENIX Association},
month = jul
}

@inproceedings{faisal2024will,
  title={When will my $\{$ML$\}$ Job finish? Toward providing Completion Time Estimates through $\{$Predictability-Centric$\}$ Scheduling},
  author={Faisal, Abdullah Bin and Martin, Noah and Bashir, Hafiz Mohsin and Lamelas, Swaminathan and Dogar, Fahad R},
  booktitle={18th USENIX Symposium on Operating Systems Design and Implementation (OSDI 24)},
  pages={487--505},
  year={2024}
}

@article{mitzenmacher2002power,
  author={Mitzenmacher, M.},
  journal={IEEE Transactions on Parallel and Distributed Systems},
  title={The power of two choices in randomized load balancing},
  year={2001},
  volume={12},
  number={10},
  pages={1094-1104},
  doi={10.1109/71.963420}}

@misc{ray,
  author = {Ray Authors},
  title = {Ray},
  url = {https://docs.ray.io/en/latest/ray-core/walkthrough.html},
  year = {2025}
}

@misc{kubernetes,
  author = {Kubernetes Authors},
  title = {Production-Grade Container Orchestration},
  url = {https://kubernetes.io/},
  year = {2025}
}

@inproceedings{gim2025pie,
  title={Pie: A Programmable Serving System for Emerging LLM Applications},
  author={Gim, In and Ma, Zhiyao and Lee, Seung-seob and Zhong, Lin},
  booktitle={Proceedings of the ACM SIGOPS 31st Symposium on Operating Systems Principles},
  pages={415--430},
  year={2025}
}

@inproceedings{lin2024parrot,
  title={Parrot: Efficient serving of $\{$LLM-based$\}$ applications with semantic variable},
  author={Lin, Chaofan and Han, Zhenhua and Zhang, Chengruidong and Yang, Yuqing and Yang, Fan and Chen, Chen and Qiu, Lili},
  booktitle={18th USENIX Symposium on Operating Systems Design and Implementation (OSDI 24)},
  pages={929--945},
  year={2024}
}

@inproceedings{zhou2024staragents,
author = {Zhou, Hang and Tang, Yehui and Qin, Haochen and Yang, Yujie and Jin, Renren and Xiong, Deyi and Han, Kai and Wang, Yunhe},
title = {Star-agents: automatic data optimization with LLM agents for instruction tuning},
year = {2024},
isbn = {9798331314385},
publisher = {Curran Associates Inc.},
address = {Red Hook, NY, USA},
booktitle = {Proceedings of the 38th International Conference on Neural Information Processing Systems},
articleno = {149},
numpages = {23},
location = {Vancouver, BC, Canada},
series = {NIPS '24}
}

@misc{mcp-atlas,
      title={MCP-Atlas: A Large-Scale Benchmark for Tool-Use Competency with Real MCP Servers},
      author={Chaithanya Bandi and Ben Hertzberg and Geobio Boo and Tejas Polakam and Jeff Da and Sami Hassaan and Manasi Sharma and Andrew Park and Ernesto Hernandez and Dan Rambado and Ivan Salazar and Rafael Cruz and Chetan Rane and Ben Levin and Brad Kenstler and Bing Liu},
      year={2026},
      eprint={2602.00933},
      archivePrefix={arXiv},
      primaryClass={cs.SE},
      url={https://arxiv.org/abs/2602.00933},
}

@misc{openclaw2026,
  title        = {{OpenClaw}: Personal AI Assistant},
  author       = {{OpenClaw Contributors}},
  year         = {2026},
  url          = {https://github.com/openclaw/openclaw},
  note         = {Accessed: 2026-05-15}
}

@misc{comfyui2026,
  title        = {{ComfyUI}: The Most Powerful and Modular Diffusion Model GUI, API and Backend with a Graph/Nodes Interface},
  author       = {{Comfy Org}},
  year         = {2026},
  url          = {https://github.com/comfy-org/ComfyUI},
  note         = {Accessed: 2026-05-15}
}

@misc{opencode2026,
  title        = {{OpenCode}: The Open Source Coding Agent},
  author       = {{Anomaly}},
  year         = {2026},
  url          = {https://github.com/anomalyco/opencode},
  note         = {Accessed: 2026-05-15}
}

@misc{codex2026,
  title        = {{Codex}: Lightweight Coding Agent that Runs in Your Terminal},
  author       = {{OpenAI}},
  year         = {2026},
  url          ={https://github.com/openai/codex},
  note         = {Accessed: 2026-05-15}
}

@dataset{dataclaw_peteromallet_2026,
  title        = {{DataClaw PeterOMallet}: Coding Agent Conversation Logs},
  author       = {{Peter O'Malley}},
  year         = {2026},
  publisher    = {Hugging Face},
  url ={https://huggingface.co/datasets/peteromallet/dataclaw-peteromallet},
  note         = {MIT License, accessed 2026-05-15}
}

@inproceedings{sglang,
author = {Zheng, Lianmin and Yin, Liangsheng and Xie, Zhiqiang and Sun, Chuyue and Huang, Jeff and Yu, Cody Hao and Cao, Shiyi and Kozyrakis, Christos and Stoica, Ion and Gonzalez, Joseph E. and Barrett, Clark and Sheng, Ying},
title = {SGLang: efficient execution of structured language model programs},
year = {2024},
isbn = {9798331314385},
publisher = {Curran Associates Inc.},
address = {Red Hook, NY, USA},
booktitle = {Proceedings of the 38th International Conference on Neural Information Processing Systems},
articleno = {2000},
numpages = {27},
location = {Vancouver, BC, Canada},
series = {NIPS '24}
}

@misc{swebench-pro,
      title={SWE-Bench Pro: Can AI Agents Solve Long-Horizon Software Engineering Tasks?},
      author={Xiang Deng and Jeff Da and Edwin Pan and Yannis Yiming He and Charles Ide and Kanak Garg and Niklas Lauffer and Andrew Park and Nitin Pasari and Chetan Rane and Karmini Sampath and Maya Krishnan and Srivatsa Kundurthy and Sean Hendryx and Zifan Wang and Vijay Bharadwaj and Jeff Holm and Raja Aluri and Chen Bo Calvin Zhang and Noah Jacobson and Bing Liu and Brad Kenstler},
      year={2025},
      eprint={2509.16941},
      archivePrefix={arXiv},
      primaryClass={cs.SE},
      url={https://arxiv.org/abs/2509.16941},
}


\end{document}